\documentclass[article]{aa}
\usepackage[varg]{txfonts}
\usepackage{graphicx}
\usepackage{natbib}
\usepackage[figuresright]{rotating}
\usepackage{hyperref}

\bibpunct{(}{)}{;}{a}{}{,} 
\def\teff{$T\rm_{eff}$}
\def\kms{$\mathrm{km\, s^{-1}}$}

\newcommand{\mygi}{MyGIsFOS}
\newcommand{\Teff}{\ensuremath{T_\mathrm{eff}}}

\newcommand{\logg}{\ensuremath{\log g}}

\newcommand{\draftflag}{false}

\newcommand{\beq}{\begin{equation}}
\newcommand{\eeq}{\end{equation}}

\newcommand{\linfor}{{\sf Linfor3D}}

\newcommand{\cfe}{{\rm [C/Fe]}}

\hypersetup{
                colorlinks, 
                citecolor=blue, 
                filecolor=blue, 
                linkcolor=blue,
                breaklinks=true, 
                plainpages=false,
                urlcolor=blue 
}

\begin{document}

\title{Investigation of a sample of carbon-enhanced metal-poor stars observed with FORS\thanks{Based on 
observations made with ESO Telescopes at the La Silla Paranal Observatory under programme ID 099.D-0791}
 and GMOS\thanks{Based on 
observations obtained at the Gemini Observatory (processed using the Gemini IRAF package), 
which is operated by the Association of Universities for Research in Astronomy, Inc., 
under a cooperative agreement with the NSF on behalf of the Gemini partnership:
the National Science Foundation (United States), the National Research Council (Canada), 
CONICYT (Chile), Ministerio de Ciencia, Tecnolog\'ia e Innovaci\'on Productiva (Argentina), 
and Minist\'erio da Ci\^encia, Tecnologia e Inova\c{c}\~{a}o (Brazil).}
}

\author{
E.~Caffau  \inst{1} \and
A.~J.~Gallagher  \inst{2,1} \and
P.~Bonifacio   \inst{1} \and
M.~Spite       \inst{1} \and
S.~Duffau      \inst{3} \and
F.~Spite       \inst{1} \and
L.~Monaco      \inst{3} \and
L.~Sbordone    \inst{4}
}

\institute{ 
GEPI, Observatoire de Paris, PSL University, CNRS,  5 Place Jules Janssen, 92190 Meudon, France
\and
Max-Planck-Institut f\"{u}r Astronomie, K\"{o}nigstuhl 17, D-69117 Heidelberg, Germany
\and
Departamento de Ciencias Fisicas, Universidad Andres Bello, Fernandez Concha 700, Las Condes, Santiago, Chile
\and
European Southern Observatory, Casilla 19001, Santiago, Chile
}
\authorrunning{Caffau}
\titlerunning{FORS/GMOS sample}
\date{Received 15/12/2017; Accepted 19/02/2018}

\abstract%
{}
{Carbon-enhanced metal-poor (CEMP) stars represent a sizeable fraction of all
known metal-poor stars in the Galaxy. Their formation and composition
remains a significant topic of investigation within the stellar
astrophysics community.}
{We analysed a sample of low-resolution spectra of 30 dwarf stars, obtained using the the visual and near UV FOcal Reducer and low dispersion Spectrograph for the Very Large Telescope (FORS/VLT) of the European Southern Observatory (ESO)
   and the Gemini Multi-Object Spectrographs (GMOS) at the GEMINI telescope, 
to derive their metallicity and carbon abundance.}
{We derived C and Ca from all spectra,  and Fe and Ba from the majority of the stars.}
{We have extended the population statistics of CEMP stars and
  have confirmed that in general, stars with a high C abundance
  belonging to the high C band show a high Ba-content (CEMP-s or
  -r/s), while stars with a normal C-abundance or that are C-rich, but
  belong to the low C band, are normal in Ba (CEMP-no).  }

 \maketitle

\section{Introduction}

The existence of stars with large carbon
enhancement was suspected from the very beginning
of astronomical spectroscopy.
In his ``Memoria'' of 1868, Angelo Secchi \citep{Secchi} added 
a fourth spectral type to the three he defined earlier and noted
that ``not all the stars of the fourth type have identical spectra: this type
allows a wider variety than the three previous ones. The black 
line after the green almost coincides with magnesium, but it may well be due to carbon. More
precise measures will decide: its width makes us believe that it is not
metallic.''\footnote{Original text in Italian: ``Non tutte le stelle del
$4^\circ$ tipo sono di spettro identico: questo tipo ammette variet\`a
maggiori che i tre precedenti.\\
La riga nera dopo il verde coincide quasi con il magnesio,
ma pu\`o bene anche appartenere al carbonio. Le misure pi\`u precise
decideranno: la sua larghezza ci fa credere che non \`e la metallica.''
}
The ``black line'' observed by Secchi  is indeed the C$_2$ Swan
band that characterises carbon stars, i.e. stars that have C/O\footnote{${\rm X/Y}=N({\rm X})/N({\rm
      Y})=10^{\left[ A({\rm X})_*-A({\rm Y})_* \right]}$}$>1$,
see \citet{McCarthy} for more details on Secchi's discovery and
a modern identification of his type 4 stars. It is remarkable
that at the low resolution of his prism spectra, Secchi was able to 
suspect that the line was too wide to be an atomic line; this proves that
Secchi was an exceptional observer.

In what can be considered a milestone review, \citet{Bidelman}
introduced the class of ``non-typical'' carbon stars, and among these
the ``CH stars'' that show ``extremely strong features due to CH,
and considerably weaker lines of neutral metals than do
the ``typical'' carbon stars.'' This was the first class of
metal-poor carbon-enhanced stars to be clearly identified, although
they are only ``mildly'' metal poor ([Fe/H]$\ga -1.5$).
As soon as it was possible to determine detailed chemical
abundances, it became clear that the composition of carbon stars
was the result of nuclear processing; this concept is well explained in the
influential review by \citet{Wallerstein}. At the time it was unclear, however,
whether this processing took place in the star itself (self-pollution) or
if it was the result of mass transfer from an evolved companion. 
By the mid-1990s, the general consensus was that Ba stars, CH-stars, and
sg-CH-stars were the result of mass transfer in a  binary system
\citep{McClure84,MW90,McClure97}. 
During its lifetime, the higher mass
star of the binary system  evolves onto the asymptotic giant branch (AGB),
where it is fuelled in part by helium burning, producing freshly synthesised
carbon and dredging it up to its atmosphere through several convective
mechanisms, induced by thermal pulsations. The giant star atmosphere 
then expands and becomes tenuous, exceeding its Roche lobe or even encompassing
the lower mass companion in the case of close binary systems. This allows
accretion of the atmosphere of the higher mass star, which  contains freshly
synthesised carbon, by the lower mass star.

The HK survey \citep{HKI,HKII} was aimed at the discovery of metal-poor stars.
One of the first surprising results was that about 10\% of the stars
with an estimated metallicity below --2.0 showed a {\it G}-band stronger than 
normal, while at higher metallicity, strong {\it G}-band stars
were less frequent \citep{N97a}. 
The first handful of strong {\it G}-band stars from the HK Survey
that were analysed at high resolution proved to be enhanced
in the $s-$process elements \citep{Barbuy,N97a}, and the exceptional
star CS\,22892-052 proved to be also enhanced in $r-$process
elements \citep{McWilliam,Sneden}. This was a strong suggestion
that these stars are indeed metal-poor analogues of CH-stars and
the result of mass-transfer in a binary system. 
The situation changed drastically with the study of CS\,22957-027
\citep{N97b,Bonifacio98}, which showed a large C-enhancement and
a very strong Swan band \citep{Bonifacio98}, but no enhancement of
any of the neutron capture elements. 
In order to distinguish these stars from the classical carbon stars 
and CH-stars, it soon become customary to refer to them as
carbon-enhanced metal-poor stars (CEMP, for short); the first two 
occurrences of the acronym in the literature are in 
\citet{Lucatello03} and in the review by \citet{Christlieb03}.

We take the definition of \cite{bc05}, and
consider  a metal-poor star as a  CEMP when $[{\rm Fe/H}]\la-2.0$\,dex and $[{\rm C/Fe}]>+1.0$\,dex.
When a CEMP star is also enhanced in 
$s-$process elements, we refer to it  as a
CEMP-s star, i.e.  $\cfe>+1.0$, ${\rm [Ba/Fe]}>+1.0$ and ${\rm [Ba/Eu]>+0.5}$ \citep{bc05}.
Another sub-class of CEMP stars show enhancements in both slow (s-) and rapid
(r-) process material, such that $\cfe>+1.0$ and $0.0<{\rm [Ba/Eu]<+0.5}$
\citep{bc05}. These stars are referred to as CEMP-r/s stars. Like CEMP-s
stars, a large fraction of CEMP-r/s stars are found in binary systems,
explaining their carbon and s-process enhancements. The r-process enhancement
is peculiar, however. It has been suggested that these stars undergo a double-enhancement phase \citep{Jonsell2006,Bisterzo2006}. The s-process and carbon
enhancement still occurs due to binary interaction, but the r-process
enhancement takes place in the forming gas cloud, when it is enriched by
r-process material. Since the discovery of GW170817 \citep{Troja2017},
observational evidence suggests that the main sites for the r-process are
neutron star mergers. The r-process-rich ejecta from these events mix with the
interstellar medium (ISM), which will become star-forming regions
and form new generations of r-rich stars.
Moreover, \citet{hampel16} could explain the pattern of the heavy
element abundances in 20 CEMP-rs stars by the accretion of the products
of the i-process \citep{cowanrose77} occuring in the intershell region
of low-metallicity AGB stars.

CEMP-s (and CEMP-r/s) stars are the most commonly found sub-class of CEMP
stars, making up approximately $80\%$ of all CEMP stars known
\citep{Aoki2007,Aoki2008}. 
It is now well established that the CEMP-s stars are all members of
binary systems
\citep{lucatello05,Starkenburg}. 
However, some CEMP stars appear to be similar
to the prototype star CS\,22957-027 and show no other
enhancements. These stars are known as CEMP-no stars \citep[$\cfe>+1.0$, ${\rm
  [Ba/Fe]}<0.0$][]{bc05}. Additionally, there is currently no evidence that
supports an unusually high fraction of binarity in these objects, but only a
handful of these objects have radial velocity measurements at this time 
\citep[e.g. the ultra Fe-poor star by][]{hydra}. 
This would suggest that accretion through binary interaction is probably not the
means by which these stars attained their carbon enhancement. Therefore, these
stars would appear to exhibit abundance patterns indicative of the gas cloud
from which they formed, making their unusual chemical composition a mystery.

\citet{spite13} suggested that CEMP stars are divided into two groups:
the high band and the low band. \citet{topos2} argued for a different
origin of the two groups, where the stars of the higher band are members of
a binary system and those that belong to the low band present the
composition of the gas cloud from where they have formed, as described
above. Further thoughts on the two C bands are discussed by
\citet{topos4}. In the case of dwarf CEMP stars, when they are
extremely metal poor, it is in general not easy to detect any sign of heavy
elements. The most Fe-poor stars known belong to the low carbon band,
while the stars in the high C band rarely have $[{\rm
    Fe/H}]\la-3.5$\,dex. 
\citet{topos4} proposed a classification for
the stars on the two C bands according their C abundance, regardless of
their abundances in neutron capture elements.

We present an analysis made on a sample of stars observed at low resolution
with the visual and near UV FOcal Reducer and low dispersion Spectrograph for the Very Large Telescope (FORS/VLT) 
of the European Southern Observatory (ESO) and the Gemini Multi-Object Spectrographs (GMOS) at the GEMINI telescope, 
in the following study, and examine, among other things, their carbon abundances.

\section{Observations and data reduction}

The spectra presented in this paper have been acquired in the course of two
programs approved by the ESO and GEMINI observatories. The programs have been
designed to be ``filler'' programs, meaning that they can acquire useful data in
weather conditions when most programs cannot operate and have no constraint on
the Moon phase. Our spectra have been acquired with FORS \citep{Appenzeller}
at the ESO VLT 8.2\,m telescope and with GMOS \citep{GMOS} at the GEMINI 8.2m
telescope.

The FORS spectra have been observed in service mode during the ESO Programme
099.D-0791, between 01/04/2017 and 16/08/2017. We used GRIS\,1200B with
a central wavelength of $436$\,nm and a slit width of 0\farcs{29}, which provides
a resolving power ${\rm R}\sim 5000$ at the central wavelength and a
continuous spectral coverage in the range $366$\,nm to $511$\,nm. The detector
used was the CCD mosaic of two $\rm 2k\times 4k$ MIT CCDs with pixel size of
$15\,\mu$m. The detectors are arranged in a line in the direction orthogonal
to the dispersion, thus there are no gaps in the spectra. This mosaic  
is more sensitive in the red; using the blue sensitive mosaic
of E2V CCDs would have been more efficient, but this detector camera is
normally not mounted on FORS. Since the program was designed as a ``filler''
program, we had to use the default detector. We initially began the
observations with a $2\times 2$ on-chip binning, that is, the default for FORS,
which is supposed to be used for service observations. We knew that in this way,
the spectrum would be under-sampled in the blue part, but our primary goal was
the {\it G}-band, which is very wide and thus would not be under-sampled; the same applies
to the \ion{Ca}{ii} K line.

We acquired spectra for eight stars in this configuration (SDSS\,J0905+0330,
SDSS\,J1248--0726, SDSS\,J1313--0019, SDSS\,J1349--0229, SDSS\,J1411+0503,
SDSS\,J2137--0057, SDSS\,J2219+0515, and SDSS\,J2239--0048). As soon as the
spectra were observed, they were reduced with the ESO FORS pipeline
\footnote{\href{ftp://ftp.eso.org/pub/dfs/pipelines/fors/fors-kit-5.3.23.tar.gz}{ftp://ftp.eso.org/pub/dfs/pipelines/fors/fors-kit-5.3.23.tar.gz}}
driven by gasgano
\footnote{\href{ftp://ftp.eso.org/pub/dfs/gasgano/gasgano-2.4.8.tar.gz}
  {ftp://ftp.eso.org/pub/dfs/gasgano/gasgano-2.4.8.tar.gz}}.  They were
subsequently analysed with \mygi\ \citep{sbordone14}. At this point, we
realised that several \ion{Fe}{i} lines could be used in the blue part of the
spectrum and that the under-sampling introduced an undesirable increase in the
uncertainty. We therefore submitted to ESO a waiver request to be allowed to
use the $1\times 1$ binning even in service mode. The waiver was approved and
all subsequent observations were acquired with a $1\times 1$ binning. For each
star we used an observing block of one hour. For the $2\times 2$ binning, this
translates into an effective exposure time of 2829\,s. For the $1\times 1$
binning, since the read-out time is longer, this translates into 2762\,s of
exposure time. Thirty spectra of quality A or B and one spectrum with quality
C have been retained for analysis for a total of 28 stars. Star
SDSS\,J144533.32--004559.0 had two ``B'' quality spectra, each of which was analysed
independently.

The GMOS spectra were acquired in service mode on the nights of 21/07/2017 and
25/07/2017. We used the B1200+\_G5321 grating centred at $468$\,nm, with a
slit of 0\farcs{5}. This combination provides a resolving power ${\rm R}=3744$
at the blaze wavelength ($463$\,nm) and a spectral coverage from $387$\,nm to
$548$\,nm. The detector was a mosaic of three Hamamatsu $\rm 2k\times 4k$ CCDs
with pixels of $15\,\mu$m side \citep{GMOS_CCD}. The three detectors are
arranged in a line along the dispersion direction, which implies that there
are two gaps in the wavelength range. The gap ranges are $438.7\,{\rm
  nm}-440.8\,{\rm nm}$ and $493.0\,{\rm nm}-495.1\,{\rm nm}$. We used a
$2\times 2$ binning, which is the default. For each star we observed three
exposures of $441$\,s. The data were processed with GEMINI
IRAF\footnote{http://iraf.noao.edu/}
package\footnote{http://www.gemini.edu/node/11823}, to perform bias
subtraction, flat-fielding, and mosaicking of the different sub-images.  We then
used the ESO MIDAS\footnote{https://www.eso.org/sci/software/esomidas/} {\tt
  LONG} context to perform the wavelength calibration using the CuAr lamp
spectra and the optimal extraction (including sky-subtraction). The three
exposures for each star where then added and analysed with \mygi.

\section{Sample}

We selected a sample of turn-off stars from the Sloan Digital Sky Survey
\citep[SDSS][]{york00,yanny09} that were bright enough ($g < 17$) to allow us to secure a
reasonable spectrum quality in a single observing block of 1\,h.  
To select turn-off stars on the SDSS, we requested $0.18\le\left(g-z\right)_0\le 0.70$
and $\left(u-g\right)_0 > 0.70$ \citep[see][]{topos1} and obtained a sample of about 20000 stars observable
from Paranal.
By examining our own analysis of the SDSS spectra, we focused
on stars for which we derived a metallicity in the range $-3.0<[{\rm Fe/H}]<-0.5$ 
and that also exhibited a strong {\it G}-band features. This meant that stars that potentially belonged to
either C bands \citep{spite13} were selected, as well as stars at the
metal-rich end of the two C bands -- ``normal'' stars, with a solar-scaled C abundance.

Table\,\ref{sample} lists the stars we examined here, along with
their coordinates, $g$-mag, and metallicities derived from Fe abundances
computed using SDSS and FORS/GMOS spectra.

\section{Analysis}

\subsection{Comparison FORS/GMOS - SDSS}

In the FORS and GMOS spectra, several \ion{Fe}{i} lines
are usually available to derive A(Fe) (and also [Fe/H]).  However, we were not
able to derive a direct Fe abundance for three stars
(SDSS\,J2219+0515, SDSS\,J1149+0723, and SDSS\,J1349-0229) because no clean Fe line was available in their spectra. Instead, we provide
an estimation of A(Fe) based on comparison of synthetic spectra and
observed on the full spectral range.  For these stars, the $\sigma$
related to the iron abundance is an estimation of the uncertainty in
this comparison.

The SDSS spectra are of lower resolution, and in this case, we derive
the metallicity, [M/H], from any available metallic feature in the
spectra, so that the metallicity is usually based on Ca, Mg, and Fe
abundances.  The synthetic spectra we used for the chemical investigation
of the SDSS spectra are enhanced in $\alpha$ elements by +0.4\,dex in the metal-poor regime, as normally expected in Galactic metal-poor stars.
So the scale provided to give the metallicity, [M/H], is related to
the Fe abundance.  This is reasonable for ``normal'' stars that
do not
show peculiar [Mg/Fe] or [Ca/Fe] ratios, which seems to be
mainly the case in this sample of stars.  In Table\,\ref{sample}
we list the
[Fe/H] derived from the FORS and GMOS spectra and the [M/H] from the
SDSS spectra.  The comparison is generally very good, but
there are some exceptions that we discuss below.  When multiple SDSS spectra of comparable quality were available, the
metallicity and the uncertainty listed in Table\,\ref{sample}
represent an average value.

Nineteen of the 27 stars in the sample present differences in the
metallicity derived from SDSS and have an iron content from the FORS and
GMOS spectra lower than 0.3\,dex, and for 22 stars, this is lower than 0.5\,dex.
For all but three spectra (SDSS\,J1349-0229, SDSS\,J2137-0057,
and SDSS\,J1411+0503), the metallicity from SDSS and [Fe/H] from
FORS/GMOS spectra agree within the uncertainties.

We present some of the more interesting facts about a small number of the stars we analysed in Table\,\ref{sample} below.

\begin{itemize}
\item SDSS\,J1349-0229 shows the largest difference in metallicity
  when the FORS and SDSS spectra are analysed. Two spectra are
  available in SDSS\,DR12, providing a large difference among
  themselves in metallicity of 0.65\,dex.  For one of them, the
  metallicity is based on a single line. We report in
  Table\,\ref{sample} the metallicity derived from the other SDSS
  spectrum, based on four features, still with a large uncertainty.
  In addition, the quality of the FORS spectrum is not good
  (classified as ``C'') and is the poorest FORS spectrum in
  our sample. The difference in the [M/H] from the best SDSS spectrum
  and [Fe/H] from the FORS spectrum is 1.44\,dex, just larger than the
  sum of the two uncertainties (1.38\,dex).  
  SDSSJ1349-0229 was also analysed by \citet{BeharaBL10} with a
temperature only 62K hotter than the value given in Table 2. We derived
from the FORS spectra the same Fe abundance as \citet{BeharaBL10}, and we
consider the results of the two analyses in very good agreement.

\item The metallicity and [Fe/H] derived from SDSS and FORS spectra,
  respectively, for SDSS\,J2137-0057 differ by 0.63\,dex, but the
  uncertainties are 0.60\,dex, so the agreement is acceptable.

\item The difference in metallicity and [Fe/H] in SDSS\,J1411+0503
  from SDSS and FORS spectra is 0.8\,dex with uncertainties of 0.62\,dex. The
  SDSS analysis is based on two spectra, providing metallicities within
  0.01\,dex, and the uncertainty is 0.4 and 0.3\,dex for the two spectra. One
  SDSS spectrum has no \ion{Ca}{ii}-K, but other features are present.  The
  0.8\,dex difference from SDSS and FORS spectra is just larger than  $1\sigma$, so we find it acceptable.

\item For the star SDSS\,J2219+0515, there is a 0.63\,dex difference between
  the metallicity derived from the SDSS spectrum and [Fe/H] derived from the
  FORS spectrum (which falls within the uncertainties of the two
  measurements) that can be explained by no-clean Fe line in the FORS
  spectrum. The A(Ca) value derived from the FORS data is in perfect
  agreement with the SDSS analysis.

\item For SDSS\,J0905+0330, the metallicity derived from the SDSS spectrum has
  a large uncertainty (1\,dex) and is based on only two features; as a
  consequence, the 0.93\,dex difference with the [Fe/H] from the FORS spectrum
  is well within $1\sigma$.

\item
 For the stars SDSS\,J0346--0558 and SDSS\,J1248--0726, Gaia \citep{gaiadr1} detected
 another source, 1.21\,mag fainter, at distances of 5\,arcsec and
 4\,arcsec, respectively.  The difference in flux from our target star
 and the other object is in both cases large enough to avoid compromising
 our analysis.

\item
SDSS\,J2011--1110: two objects are detected by Gaia within 2\,arcsec,
with a difference in the Gaia mag of 2.53.  The two objects are very
close, but the fainter star is too faint to interfere with the signal
from our target.

\item
SDSS\,J1315+1727, SDSS\,J1337+0836, SDSS\,J1337+0836,
SDSS\,J1337+0837, and SDSS\,J1343+0810 have a 2MASS identification.

\end{itemize}

\begin{table*}
{\small
  \caption{Sample of stars. [M/H] from SDSS is based on several
    metallic features; [Fe/H] from FORS and GMOS spectra are taken from
    \ion{Fe}{i} lines only.  The solar A(Fe)=7.52 adopted is taken from
    \citet{abbosun}. The
    $\sigma$ are line-to-line scatter, and for the
    uncertain cases, the colon in the iron abundance is an estimate of
    the uncertainty in the comparison to synthetic spectra.}
\label{sample}
\renewcommand{\tabcolsep}{2pt}
\tabskip=1pt
\begin{center}
\begin{tabular}{lrrrrrrrrrrrrr}
\hline\hline
\noalign{\smallskip}
\multicolumn{1}{c}{Star}   & \multicolumn{1}{c}{RA}       & \multicolumn{1}{c}{Dec}          & \multicolumn{1}{c}{$g^a$}
 & \multicolumn{1}{c}{MJD}  & \multicolumn{1}{c}{VR}   &\multicolumn{1}{c}{$\sigma$}& \multicolumn{1}{c}{VR}       & 
\multicolumn{1}{c}{$\sigma$} & \multicolumn{1}{c}{[M/H]} & \multicolumn{1}{c}{$\sigma$} & \multicolumn{1}{c}{[Fe/H]} & \multicolumn{1}{c}{$\sigma$}\\
                           & \multicolumn{1}{c}{$^\circ$} & \multicolumn{1}{c}{$^\circ$}     &     &      & \multicolumn{1}{c}{{\tiny  SDSS}} &
  &          &          & \multicolumn{1}{c}{{\tiny SDSS}}  &          &   & \\
\hline \noalign{\smallskip}                                                                           
\multicolumn{12}{c}{FORS}\\
\hline \noalign{\smallskip}                                                                           
SDSS\,J001547.45+001326.6  & 3.9477144 & +0.22406663 & 15.81  & 57953.33251196 & $-20$   & 2$^a$& $-10$   & 20 & $-0.75$ & 0.24 & $-0.70$ & 0.17 \\ 
SDSS\,J004036.97+002540.7  & 10.154042 & +0.42797715 & 16.29  & 57953.37109608 & $+82  $ & 3    & $+92$   & 19 & $-1.99$ & 0.22 & $-1.99$ & 0.21 \\ 
SDSS\,J004252.51+005521.8  & 10.718813 & +0.92273976 & 16.83  & 57956.28170347 & $-256 $ & 4    & $-270$  & 29 & $-1.70$ & 0.24 & $-1.65$ & 0.13 \\ 
SDSS\,J022226.20-031338.0  & 35.609189 & --3.2272445 & 16.95  & 57956.32784048 & $-124 $ & 3    & $-90$   & 32 & $-2.22$ & 0.49 & $-2.63$ & 0.24 \\ 
SDSS\,J030929.93+054246.4  & 47.374692 & +5.7129120  & 15.25  & 57966.32655116 & $-82  $ & 2    & $-57$   & 35 & $-0.96$ & 0.26 & $-0.96$ & 0.16 \\ 
SDSS\,J034635.17--055818.3 & 56.64657  & --5.9717720 & 16.96  & 57973.36030252 & $+19  $ & 2    & $+32$   & 27 & $-1.20$ & 0.26 & $-0.85$ & 0.19 \\ 
SDSS\,J090536.50+033034.5  & 136.40208 & +3.5096110  & 16.86  & 57852.99994682 & $+272 $ & 4    & $+295$  &21  & $-2.19$ & 1.03 & $-3.12$ & 0.21 \\ 
SDSS\,J114932.51+072347.0  & 177.38547 & +7.3963985  & 16.34  & 57953.98955476 & $-15  $ & 3    & $-33$   &34  & $-2.88$ & 0.15 & $-2.82^b$:& 0.40 \\
SDSS\,J124841.09--072646.6 & 192.17123 & --7.4462930 & 16.84  & 57845.30650769 & $-7   $ & 3    & $-45$   & 26 & $-1.28$ & 0.45 & $-1.64$ & 0.39 \\ 
SDSS\,J131550.68+172707.0  & 198.96119 & +17.451962  & 16.12  & 57940.06317480 & $+306 $ & 3    & $+315$  & 34 & $-2.22$ & 0.67 & $-2.52$ & 0.06 \\ 
SDSS\,J133704.70+083523.4  & 204.26961 & +8.5898340  & 16.06  & 57951.97732700 & $-21  $ & 3    & $-19$   & 23 & $-1.38$ & 0.15 & $-1.31$ & 0.17 \\ 
SDSS\,J133750.48+083610.8  & 204.46034 & +8.6030160  & 16.72  & 57952.02268752 & $+7   $ & 3    & $+16$   & 28 & $-0.99$ & 0.37 & $-0.78$ & 0.18 \\ 
SDSS\,J133753.36+083734.2  & 204.47237 & +8.6261760  & 15.78  & 57955.06167568 & $-34  $ & 3    & $-11$   & 34 & $-0.51$ & 0.26 & $-0.45$ & 0.25 \\  
SDSS\,J133802.22+083056.2  & 204.50927 & +8.5156120  & 16.93  & 57956.00287188 & $-19  $ & 3    & $-10$   & 27 & $-0.51$ & 0.51 & $-0.53$ & 0.18 \\  
SDSS\,J134247.69+083521.0  & 205.69872 & +8.5891910  & 15.54  & 57981.98808043 & $-42  $ & 4    & $-23$   & 31 & $-1.63$ & 0.18 & $-1.48$ & 0.23 \\ 
SDSS\,J134343.08+081029.3  & 205.92952 & +8.1748110  & 14.92  & 57981.98808043 & $-13  $ & 2    & $-4 $   & 34 & $-0.47$ & 0.36 & $-0.50$ & 0.22 \\ 
SDSS\,J134913.54--022942.8 & 207.30642 & --2.4952258 & 16.64  & 57916.14995942 & $+140 $ & 3    & $+128$  &27  & $-1.58$ & 0.98 & $-3.02^b$:& 0.40 \\ 
SDSS\,J141123.09+050345.6  & 212.84622 & +5.0626727  & 16.03  & 57933.97656561 & $-238 $ & 4    & $-210$  &25  & $-2.56$ & 0.35 & $-3.36$ & 0.27 \\ 
SDSS\,J144533.32--004559.0 & 221.38887 & --0.7663930 & 15.32  & 57941.10539190 & $-9   $ & 2    &$-13$    & 30 & $-3.04$ & 0.22 & $-2.94$ & 0.19 \\ 
SDSS\,J144533.32--004559.0 & 221.38887 & --0.7663930 & 15.32  & 57941.13882949 & $-9   $ & 2    &$-19$    & 30 & $-3.04$ & 0.22 & $-2.94$ & 0.19 \\ 
SDSS\,J154338.58+092904.9  & 235.91078 & +9.4847057  & 16.04  & 57952.06105676 & $+48  $ & 2    & $+53$   &23  & $-2.41$ &      & $-2.38$ & 0.23 \\ 
SDSS\,J160646.36+052218.2  & 241.69318 & +5.3717390  & 16.88  & 57954.10306751 & $-85  $ & 3    & $-97$   & 22 & $-1.02$ & 0.44 & $-1.27$ & 0.29 \\ 
SDSS\,J213752.52--005754.3 & 324.46883 & --0.96510684& 16.74  & 57915.30185193 & $-108 $ & 5    & $-73$   &23  & $-2.46$ & 0.52 & $-3.09$ & 0.08 \\ 
SDSS\,J221911.25+051519.5  & 334.79691 & +5.2554223  & 16.94  & 57915.34479028 & $-232 $ & 3    & $-259$  &24  & $-2.85$ & 0.32 & $-2.22^b$:& 0.40 \\ 
SDSS\,J223946.42--004827.7 & 339.94342 & --0.8077150 & 15.73  & 57938.28656970 & $-145 $ & 2    & $-149$  & 22 & $-1.24$ & 0.45 & $-1.49$ & 0.22 \\ 
SDSS\,J231108.61+002650.3  & 347.78588 & +0.44731512 & 16.77  & 57940.32010047 & $-64  $ & 3    & $-60$   & 21 & $-1.46$ & 0.32 & $-1.68$ & 0.27 \\ 
SDSS\,J233526.49+081905.9  & 353.86042 & +8.3183210  & 16.53  & 57940.35885623 & $+32  $ & 3    & $+48$   & 23 & $-1.35$ & 0.12 & $-1.22$ & 0.19 \\ 
SDSS\,J233757.06+143607.2  & 354.48775 & +14.602015  & 16.91  & 57951.22060599 & $-49  $ & 2    & $-43$   & 38 & $-0.81$ & 0.23 & $-0.85$ & 0.23 \\ 
SDSS\,J131326.89--001941.4 & 198.362066& --0.32766   & 16.87  & 57919.0637425  & \\ 
\hline \noalign{\smallskip}                                                                           
\multicolumn{12}{c}{GMOS}\\
\hline \noalign{\smallskip}                                                                           
SDSS\,J111434.26--120214.0 & 168.64275 & --12.0372389& 16.63  & 57955.174264249& $-4   $ & 2    &         &    & $-0.84$ & 0.22 & $-1.07$ & 0.22 \\
SDSS\,J201114.15--111003.9 & 302.8089583&--11.16775  & 16.31  & 57955.147332606& $-102 $ & 2    &         &    & $-1.22$ & 0.30 & $-1.03$ & 0.17 \\
SDSS\,J212351.59--080416.3 & 320.9649583& --8.0712167& 16.73  & 57959.973129110& $-334 $ & 4    &         &    & $-2.89$ & 0.09 & $-2.91$ & 0.09 \\
\hline \noalign{\smallskip}
\noalign{\smallskip}
\hline
\end{tabular}
\end{center}
$^a$  $g$ magnitudes are those used to derive the effective
temperatures and are taken from the SDSS-DR\,12.\\ 
$^b$ estimation from comparison to synthetic spectra; we consider these measurements to be uncertain.}
\end{table*}

\subsection{Radial velocities}

The radial velocities have been derived with the cross-correlation of
each spectrum with a synthetic spectrum with very similar parameters,
also taking into account the C enhancement of the star.  We are aware
that FORS suffers from the telescope flexions, which makes the  radial velocity measurements uncertain.  We took into account
the position of the stars in the sky at the time of the observation,
and using the FORS manual indications, we derived the uncertainties
in radial velocities. These values are higher than the
formal uncertainty in each cross-correlation, and for the majority of
the spectra, taking it into account does not change the total
uncertainty.  For three spectra (for the stars
SDSS\,J2239--0048, SDSS\,J1248--0726, and SDSS\,J0905+0330), the
formal uncertainty increases the total uncertainty by 1\,\kms\ , and
SDSS\,J1411+0503, for which the quality of the cross-correlation is
worse, still shows a clear peak, and the total uncertainty is
increased by 3\,\kms.  For the GMOS stars we had no information on
the telescope flexions, so that at present we prefer to not provide the
radial velocities.  Comments on some stars are provided below.

\begin{itemize}

\item SDSS\,J0015+0013 has six SDSS spectra. The metallicities are in
  very good agreement, with a scatter of 0.06\,dex.  Of the six radial
  velocities, five are compatible within the uncertainties, but one is not
  ($-18\pm 2$, $-24\pm 3$, $-20\pm 3$, $-8\pm 3$, $-19\pm 2$, and $-18\pm
  2$). The star is probably a member of a multiple system.
  Our value of $-10\pm 20$ lies within the SDSS values, but the
  uncertainty is large enough to be compatible with any SDSS value.

\item SDSS\,J0346-0558 has three SDSS spectra. Two of them are very similar in
  metallicity (the average value is listed in Table\,\ref{sample}) and agrees within
  the uncertainties with the FORS analysis, while the third failed
  the chemical analysis.  This failure of this latest spectrum is related to
  the peculiar value of the radial velocity of --1159\,\kms\ that is provided.  The other
  two SDSS spectra provide very similar radial velocities values, well within
  the SDSS uncertainties, and the average value is provided in Table\,\ref{sample}.

\item
For SDSS\,J1606+0522, we find a double peak in the cross-correlation at
  77\,\kms \ from the mean peak.

\item The difference between the SDSS and the FORS radial velocities
  in SDSS\,J1248--0726 is larger than the uncertainties.  This can be
  an indication of a multiple system.

\item SDSS\,J2335+0819 has two SDSS spectra whose radial velocities are in
  agreement well within the uncertainties. The metallicity derived from the
  spectra agrees within the uncertainties, but for one of them, the metallicity is
  based on many more features, with a much smaller uncertainty; we list
  this result in Table\,\ref{sample}.

\item For SDSS\,J2239-0048, three SDSS spectra are available. The
  metallicities derived from these spectra are in perfect agreement (the average value is provided in
  Table\,\ref{sample}). The radial velocities
  indicate an object in a multiple system; the three values ($-128\pm
  4$, $-141\pm 3,$ and $-145\pm 2$) are incompatible with a constant radial
  velocity.  The FORS radial velocity is compatible with any SDSS value.

\item The two spectra of SDSS\,J0042+0055 in SDSS give metallicities and
  radial velocities in very good agreement.  The metallicity is in perfect
  agreement with [Fe/H] derived from the FORS spectrum, and the radial velocity
  difference of 14\,\kms\ is well within the uncertainties.

\item The disagreement between the SDSS and FORS radial velocities for
  SDSS\,J1411+0503 and SDSS\,J2137--0057 is just larger than the
  uncertainties.

\item Twelve spectra of SDSS\,J0040+0025 are available in SDSS. The
  metallicities we derived from them are in excellent agreement ($\sigma =
  0.04$\,dex) and also agree excellently with the [Fe/H] derived
  from the FORS spectrum.  The 12 radial velocities span from 75 to
  86\,\kms. The scatter on them is 3\,\kms, which is compatible with the
  uncertainties, but might indicate a variation in the radial velocity of this
  star. The radial velocity derived from the FORS spectrum is compatible with
  the values from the SDSS spectra.

\item The radial velocities of the two SDSS spectra of SDSS\,J1315+1727 agree well and lie well within the uncertainties. They also agree within the uncertainties
  with the measurement from the FORS spectrum. The two metallicities also agree well and agree with the [Fe/H] from FORS spectrum within
  the uncertainties.

\item The two FORS spectra of SDSS\,J1445-0045 have radial velocities and
  [Fe/H] in agreement within the uncertainties, and they both agree with the
  values from the SDSS spectrum.

\item The two SDSS spectra of SDSS\,J2137-0057 provide extremely good
  agreement in the metallicities derived and with [Fe/H] derived from the GMOS
  spectrum, while the radial velocities of $-95\pm 5$ and $-108\pm 5$ could
  indicate a variation (the highest value is reported in Table\,\ref{sample}).
  We were unable to derive a robust radial velocity from the GMOS spectrum.

\item The two spectra of SDSS\,J1349-0229 in SDSS show a
  difference in radial velocities that is larger than the uncertainties ($140\pm 3$
  and $131\pm 3$, the first value is listed in Table\,\ref{sample}). The value from the
  FORS spectrum is compatible with both values.

\item The radial velocities derived from the two SDSS spectra of
  SDSS\,J1114--1202 are consistent within the uncertainties. We were unable to
  derive a reliable radial velocity from the GMOS spectrum. Metallicity and
  [Fe/H] from the SDSS and GMOS spectra agree within uncertainties.

\item The two SDSS spectra of SDSS\,J2011--1110 provide radial velocities
  that are perfectly consistent.  Metallicity and [Fe/H] from the SDSS and
  GMOS spectra agree within uncertainties.

\end{itemize}

\subsection{Chemical analysis}

The stars were selected as turn-off stars, so that we fixed the
surface gravity at 4.0 in the analysis. The temperature was derived from the Sloan colours as
in \citet{topos1}. To derive the chemical composition, we analysed the spectra
with the pipeline \mygi\ \citep{sbordone14}. The grid of theoretical syntheses
were computed with {\tt Turbospectrum} \citep{alvarez_plez,turbo} on a
MARCS grid of models \citep{G2008}.

\begin{figure*}
\begin{center}
\resizebox{\hsize}{!}{\includegraphics[draft = \draftflag, clip=true]{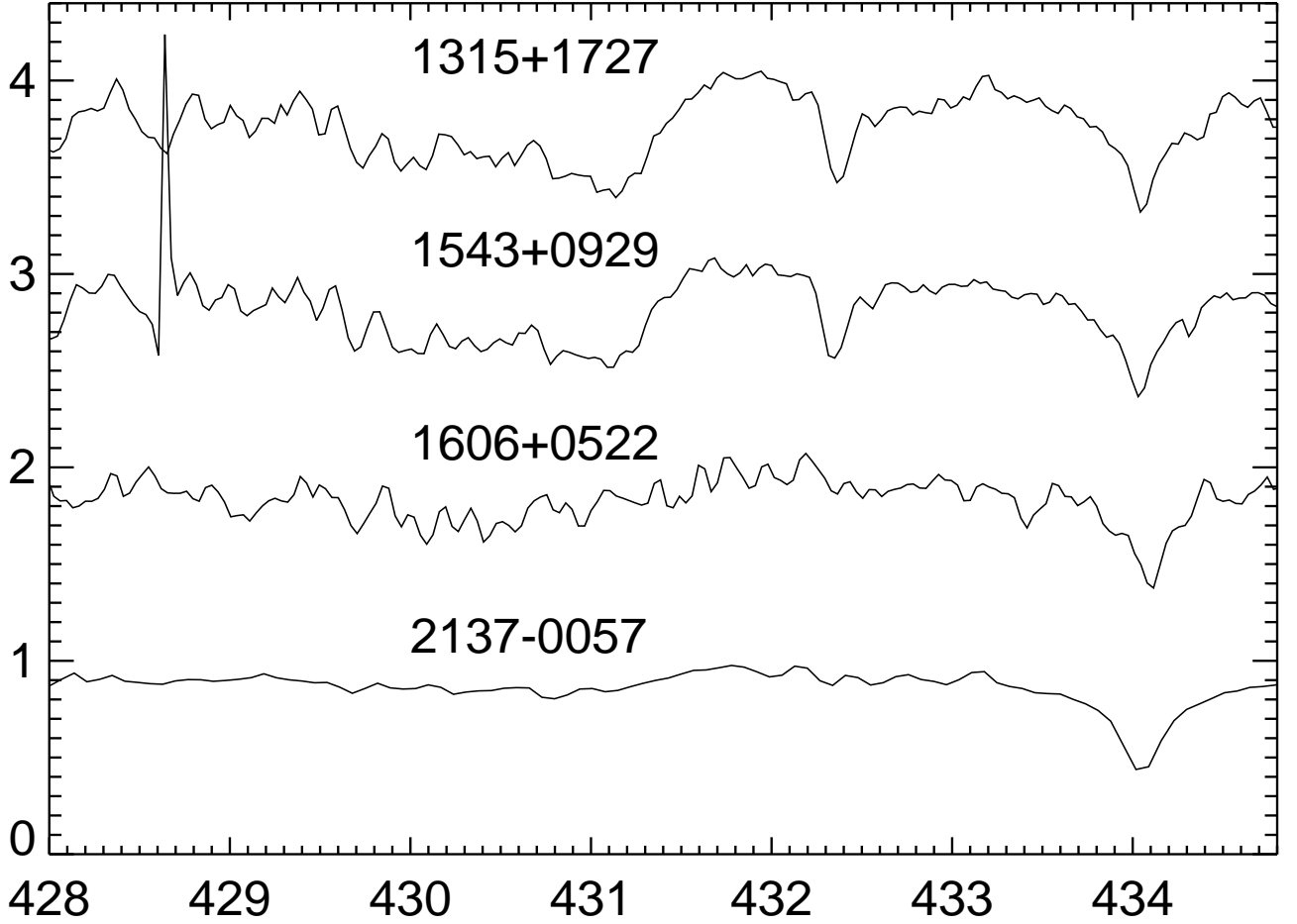}}
\end{center}
\caption[]{Observed spectra for four stars in the range of the {\it G}-band.}
\label{obs}
\end{figure*}

\begin{figure}
\begin{center}
\resizebox{\hsize}{!}{\includegraphics[draft = \draftflag, clip=true]{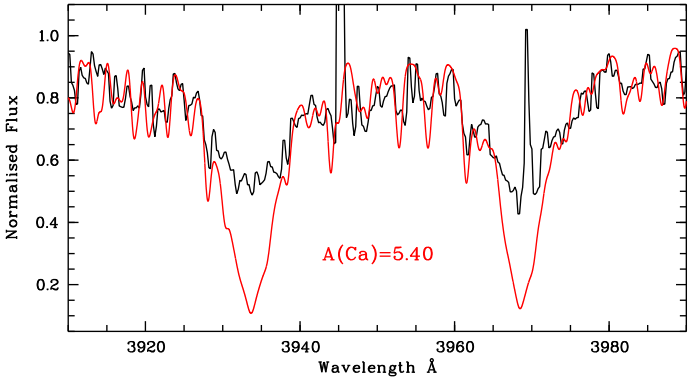}}
\end{center}
\caption[]{Observed spectrum of SDSS\,J1114--1202 (solid black) compared
  with a synthetic spectrum (solid red) with A(Ca) compatible with the value
  derived from the other two \ion{Ca}{i} lines and with A(Fe).}
\label{ca1114m1202}
\end{figure}

After we determined A(Fe), we produced a grid of synthetic spectra
with the same Fe content, but various C abundances and fitted the {\it G}-band
in order to derive A(C). We repeated this to derive A(Ca) from the
\ion{Ca}{ii}-K line and the Sr and Ba abundances. Because of the low
resolution of the spectra, we were only able to derive a few elements, but
because the \ion{Ca}{ii}-K line and {\it G}-band are so prominent, we were able to
derive Ca and C from the {\it G}-band from any spectrum. For many spectra,
we were also able to derive the abundances of Fe, Ti, Mn, Sr, and Ba. The
detailed abundances are reported in Table\,\ref{abbo}, and in
  Fig.\,\ref{obs}, we show four of the spectra.
We have an uncertainty of 0.15\,dex in deriving A(Ca) from
the \ion{Ca}{ii}-K line. The A(Ca) derived from the \ion{Ca}{ii}-K
line usually agrees within uncertainties with the value derived from
\ion{Ca}{i} lines, except for SDSS\,J1606+0522, SDSS\,J0905+0330,
and SDSS\,J1114--1202. For the star SDSS\,J1606+0522, an uncertainty of
0.31\,dex is reasonable because of the spectrum quality, but the scatter
of 0.01\,dex among the two \ion{Ca}{i} lines is a  random chance event.
For the star SDSS\,J0905+0330, the A(Ca) is based on the single
\ion{Ca}{i} line at 422\,nm, which is often in disagreement with the
other lines. The star is metal poor, therefore a contamination from the ISM
of the \ion{Ca}{ii}-K could increase the A(Ca) derived from this
line. For the star SDSS\,J1114--1202, the \ion{Ca}{ii}-K line was
clearly too shallow.  In Fig.\,\ref{ca1114m1202} we compare the
\ion{Ca}{ii}-K line with a synthetic spectrum with a Ca abundance
compatible with A(Ca) derived from two \ion{Ca}{i} lines and
consistent with the Fe abundance.  The wings of the
\ion{Ca}{ii}-H and -K lines in the synthesis reproduce the observed
spectrum very well, while the core of the line is not present. We
suspect that this is due to chromospheric emission. The H-lines in the
spectrum of this star are also weak.

In Table\,\ref{abbo} we also provide a 3D correction on the C abundance as
derived from the {\it G}-band by \citet{ajg16}. The solar A(Fe) used to produce the
figures is adopted from \citet{abbosun}, while the other solar abundances are taken from
\citet{lodders09}.

\begin{table*}
{\small
\caption{Derived abundances.}
\label{abbo}
\renewcommand{\tabcolsep}{2pt}
\tabskip=0pt
\begin{center}
\begin{tabular}{lllllllllllrr}
\hline\hline
\noalign{\smallskip}
\multicolumn{1}{c}{Star}    & \multicolumn{1}{c}{\Teff} & \multicolumn{1}{c}{A(Fe)}          & 
\multicolumn{1}{c}{A(FeII)} &  \multicolumn{1}{c}{A(C)} & \multicolumn{1}{c}{3Dcor} &    \multicolumn{1}{c}{A(Mg)} &  
\multicolumn{1}{c}{A(Si)}   & \multicolumn{1}{c}{A(CaK)/A(Ca)}         &         \multicolumn{1}{c}{A(Ti)}  &          
\multicolumn{1}{c}{A(Mn)}   &   \multicolumn{1}{c}{A(Sr)}  & \multicolumn{1}{c}{A(Ba)} \\
\hline \noalign{\smallskip}                                          
SDSSJ0015+0013    & 6065 & $6.82\pm 0.17$  & $7.03\pm 0.18$ & 7.82  & $-0.03$ &     7.28 &   7.22   & 5.81/$6.01\pm 0.21$ & $4.45\pm 0.20$ & $4.79\pm 0.21$ &        &  1.74 \\
SDSSJ0040+0025    & 5651 & $5.53\pm 0.21$  &                & 6.53  & $-0.15$ &     6.02 &          & 4.76/$4.64\pm 0.18$ & $3.14\pm 0.27$ & $3.62\pm 0.24$ &  0.46  &  0.54 \\
SDSSJ0042+0055    & 6019 & $5.87\pm 0.13$  &                & 6.79  & $-0.15$ &     6.34 &          & 5.21/$5.03\pm 0.15$ & $3.75\pm 0.17$ & $3.77\pm 0.08$ &  1.67  &  1.35 \\
SDSSJ0222--0313   & 6345 & $4.89\pm 0.24$  &                & 8.08  & $-0.08$ &          &          & 4.11/$4.12\pm 0.30$ & $3.17\pm 0.01$ &                &  2.54  &  1.52 \\
SDSSJ0309+0542    & 6012 & $6.56\pm 0.16$  & $6.36\pm 0.20$ & 7.74  & $+0.05$ &     7.17 &          & 5.71/$5.73\pm 0.22$ & $4.42\pm 0.32$ & $4.49\pm 0.14$ &  2.02  &  1.15 \\
SDSSJ0346--0558   & 5660 & $6.67\pm 0.19$  & $6.51\pm 0.19$ & 7.61  & $+0.10$ &     7.06 &          & 5.62/$5.55\pm 0.06$ & $4.27\pm 0.25$ & $4.74\pm 0.30$ &  2.00: &  1.50 \\
SDSSJ0905+0330    & 6263 & $4.40\pm 0.27$  &                & 7.68  & $-0.15$ &          &          & 3.99/$3.52\pm 0.30$ &                &                &  0.40: &  1.47:\\
SDSSJ1149+0723    & 6000 & $4.7 \pm 0.40$: &                & 7.66  & $-0.05$ &          &          & 3.86/$3.70\pm 0.30$ &                & $2.50\pm 0.30$ &$<1.0$  &  0.30: \\
SDSSJ1248--0726   & 6234 & $5.88\pm 0.39$  &                & 7.94  & $-0.10$ &     6.40 &          & 4.99/$5.09\pm 0.32$ & $3.62\pm 0.13$ & $4.09\pm 0.30$ &  2.64  &  3.05 \\
SDSSJ1315+1727    & 6170 & $5.00\pm 0.06$  &                & 8.34  & $-0.05$ &          &          & 3.99                &                &                &  0.30: &  1.81 \\
SDSSJ1337+0835    & 6023 & $6.21\pm 0.17$  &                & 7.35  & $+0.00$ &     6.91 &          & 5.38/$5.39\pm 0.13$ & $3.74\pm 0.28$ & $4.19\pm 0.18$ &  1.80  &  0.90 \\
SDSSJ1337+0836    & 5970 & $6.74\pm 0.18$  &                & 8.44  & $+0.15$ &     7.18 &          & 6.27/$6.00\pm 0.38$ & $4.32\pm 0.18$ & $4.39\pm 0.30$ &        &  1.90 \\
SDSSJ1337+0837    & 5464 & $7.07\pm 0.25$  &                & 8.10  & $+0.15$ &          &          & 6.09/$6.22\pm 0.24$ & $4.46\pm 0.28$ & $4.88\pm 0.31$ &        &  1.82 \\
SDSSJ1338+0830    & 5496 & $6.99\pm 0.18$  &                & 8.04  & $+0.15$ &     7.61 &          & 6.02/$6.10\pm 0.26$ & $4.72\pm 0.23$ & $4.94\pm 0.24$ &        &  1.67:\\
SDSSJ1342+0835    & 6176 & $6.04\pm 0.23$  &                & 7.00: & $-0.10$ &          &          & 5.21/$4.91\pm 0.16$ & $3.89\pm 0.24$ & $3.68\pm 0.30$ &  1.85  &  0.51 \\
SDSSJ1343+0810    & 5482 & $7.02\pm 0.22$  &                & 7.97  & $+0.15$ &     7.45 &   7.37   & 6.03/$6.02\pm 0.07$ & $4.72\pm 0.16$ & $5.33\pm 0.14$ &        &  1.72 \\
SDSSJ1349--0229   & 6138 & $4.5 \pm 0.40$: &                & 8.32  & $+0.00$ &          &          & 3.78                &                &                &  1.00: &  1.50 \\
SDSSJ1411+0503    & 5930 & $4.16\pm 0.23$  &                & 6.86  & $-0.15$ &          &          & 3.91                &                &                & $-0.19$&$<-0.8$\\ 
SDSSJ1445--0045,2 & 5492 & $4.58\pm 0.19$  &                & 6.18  & $-0.15$ &          &          & 3.64/$3.30\pm 0.30$ & $2.65\pm 0.26$ &                & $-0.12$&$<-1.0$ \\
SDSSJ1445--0045,1 & 5492 & $4.49\pm 0.21$  &                & 6.18  & $-0.15$ &          &          & 3.63/$3.49\pm 0.29$ & $2.53\pm 0.06$ &                & $-0.34$&$<-1.0$ \\
SDSSJ1543+0929    & 6288 & $5.14\pm 0.08$  &                & 8.42  & $-0.05$ &          &          & 4.05                &                &                &  1.13: &  1.21: \\
SDSSJ1606+0522    & 6122 & $6.25\pm 0.29$  & $6.55$         & 7.81  & $+0.00$ &     6.58 &          & 5.38/$5.69\pm 0.01$ & $4.59\pm 0.04$ &                &  3.29  &  2.95 \\
SDSSJ2137--0057   & 5956 & $4.43\pm 0.40$  &                & 6.70  & $-0.15$ &          &          & 3.73                &                &                &$<-0.5$ & --0.02 \\
SDSSJ2219+0515    & 6351 & $5.3 \pm 0.40$: &                & 8.36  & $+0.00$ &          &          & 4.11                &                &                &$<0.0$  &  1.40:\\
SDSSJ2239--0048   & 6317 & $6.03\pm 0.22$  &                & 7.91  & $-0.05$ &     6.46 &          & 5.19/$5.07        $ & $3.94\pm 0.39$ &                &  3.31  &  3.45 \\
SDSSJ2311+0026    & 5967 & $5.84\pm 0.27$  &                & 7.06  & $-0.10$ &     6.70 &          & 5.20/$5.03\pm 0.04$ & $3.67\pm 0.07$ & $3.72\pm 0.30$ &  1.77  &  1.10 \\
SDSSJ2335+0819    & 5951 & $6.30\pm 0.19$  &                & 7.36  & $+0.05$ &     6.93 &          & 5.42/$5.32\pm 0.05$ & $4.23\pm 0.36$ & $4.39\pm 0.38$ &  1.70  &  0.88 \\
SDSSJ2337+1436    & 5746 & $6.67\pm 0.23$  &                & 7.76  & $+0.10$ &     7.34 &          & 5.80/$5.79\pm 0.18$ & $4.14\pm 0.34$ & $4.39\pm 0.30$ &        &  1.45 \\
SDSSJ1114--1202   & 5879 & $6.45\pm 0.22$  &                & 7.61  & $+0.10$ &$7.18\pm 0.22$ &     &5.15:/$5.47\pm 0.11$ & $3.87\pm 0.09$ &                &  1.00: & 0.92: \\
SDSSJ2011--1110   & 5540 & $6.49\pm 0.17$  & $6.66\pm 0.07$ & 7.35  & $+0.10$ &$7.18\pm 0.11$ & 7.13&5.36:/$5.45\pm 0.09$ & $4.05\pm 0.21$ & $4.45\pm 0.42$ &  1.84: &  1.35 \\
SDSSJ2123--0804   & 5468 & $4.61\pm 0.09$  &                & 6.00  & $-0.10$ & $5.00\pm 0.13$ &     & 3.91/$3.74\pm 0.34$ & $2.41$         &                &  0.47: &  0.46 \\
\hline \noalign{\smallskip}
\noalign{\smallskip}
\hline
\end{tabular}
\end{center}
A colon indicates uncertain values.
}
\end{table*}

\begin{figure}
\begin{center}
\resizebox{\hsize}{!}{\includegraphics[draft = \draftflag, clip=true]{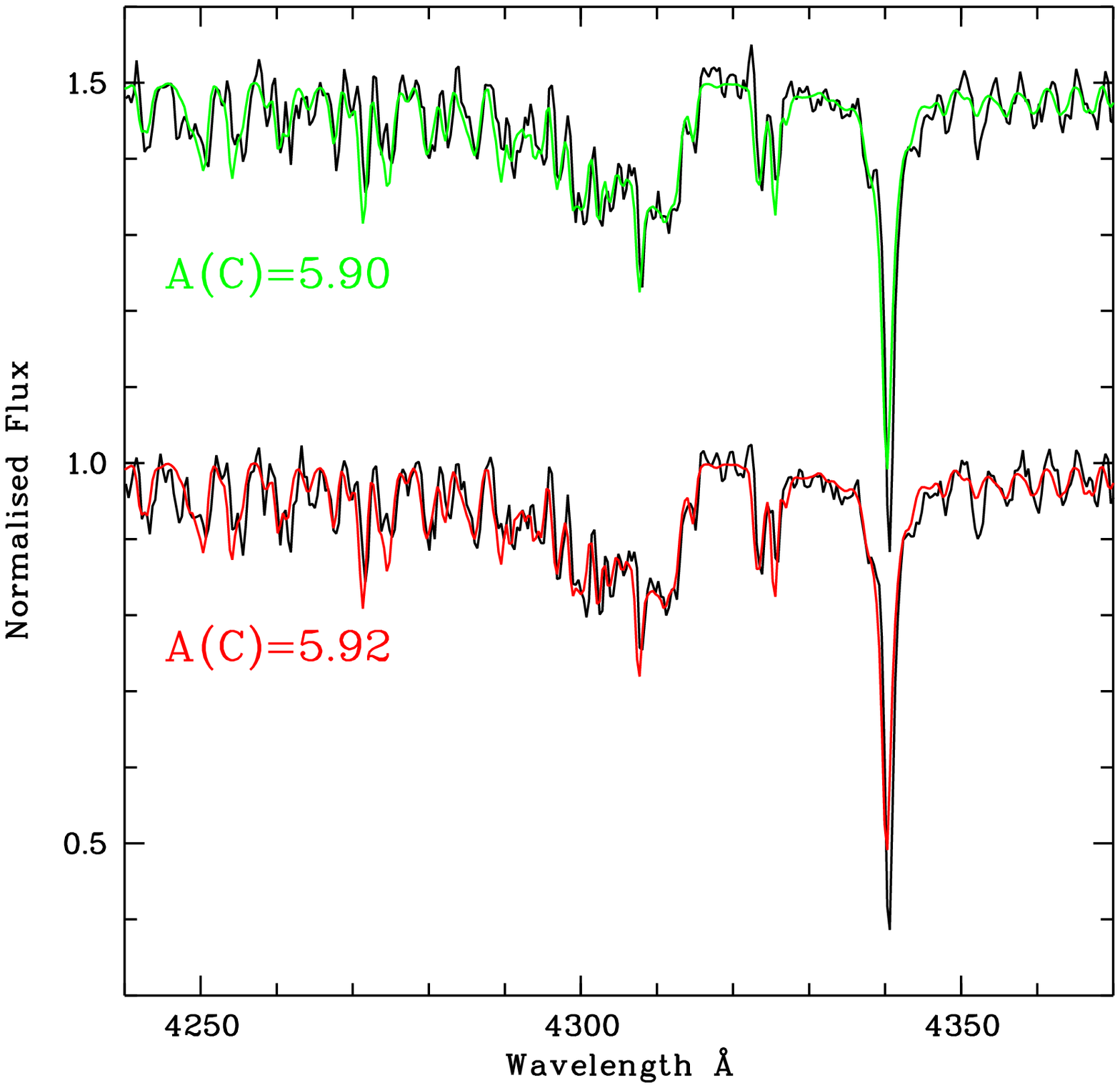}}
\end{center}
\caption[]{Two observed spectra of SDSS\,J1445--0045 (solid black) compared
  with the best fit (solid green and red) based on a 3D synthesis of a model
that is  18\,K cooler.}
\label{g1445m0045}
\end{figure}

\begin{figure}
\begin{center}
\resizebox{\hsize}{!}{\includegraphics[draft = \draftflag, clip=true]{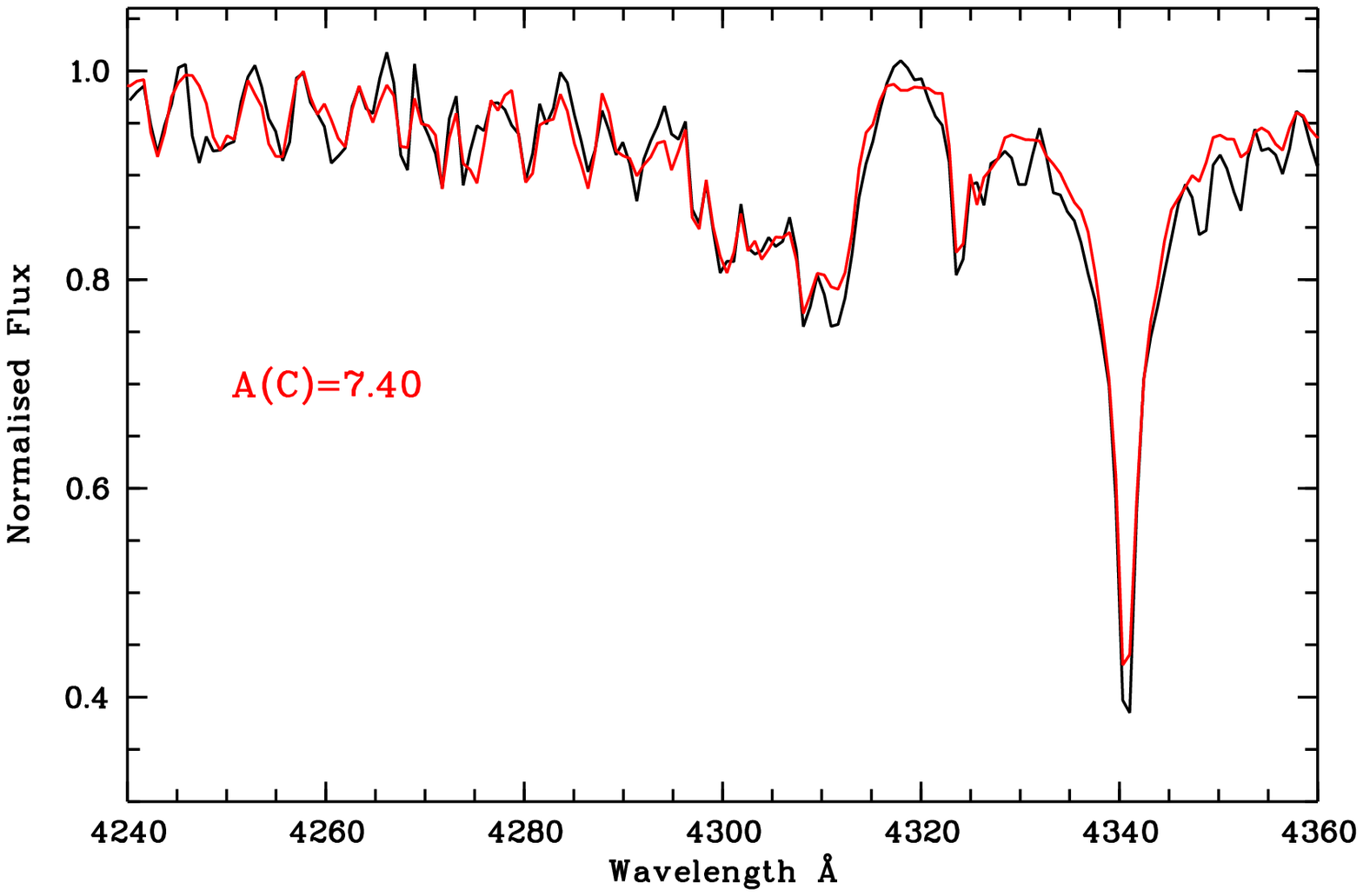}}
\end{center}
\caption[]{Observed spectrum of SDSS\,J0905+0330 (solid black) compared
  with the best fit (solid red) based on a 3D synthesis of a model that is 21\,K
  cooler.}
\label{g0905p0330}
\end{figure}

For two stars, SDSS\,J1445--0045 and SDSS\,J0905+0330, we have
hydrodynamical models from the CIFIST grid \citep{ludwig09} with
  parameters very similar to what is observed. We fitted the G
bands
  of these two stars with the synthetic spectra computed using these
  3D models with
  \linfor\footnote{\href{http://www.aip.de/Members/msteffen/linfor3d}{http://www.aip.de/Members/msteffen/linfor3d}}
  \citep{ajg17}.  The best fits are presented in
Fig.\,\ref{g1445m0045} and \ref{g0905p0330}.  As was explained in
\citet{ajg16}, the {\it G}-band is extremely sensitive to changes in the
oxygen abundance. Because we were unable to directly measure the oxygen
abundance, the 3D syntheses were computed so that oxygen scales
with carbon at a consistent ratio corresponding to A(C)--A(O) of
--0.67.

\subsection{SDSS\,J1313--0019}
In the sample of stars, we also observed SDSS\,J1313--0019, a star
discovered by \citet{carlos15} that was later also analysed by
\citet{frebel15}.  Both analyses show that the star is an
evolved star, extremely low in [Fe/H], with a C content that places the
star in the low C band.

\begin{figure}
\begin{center}
\resizebox{\hsize}{!}{\includegraphics[draft = \draftflag, clip=true]{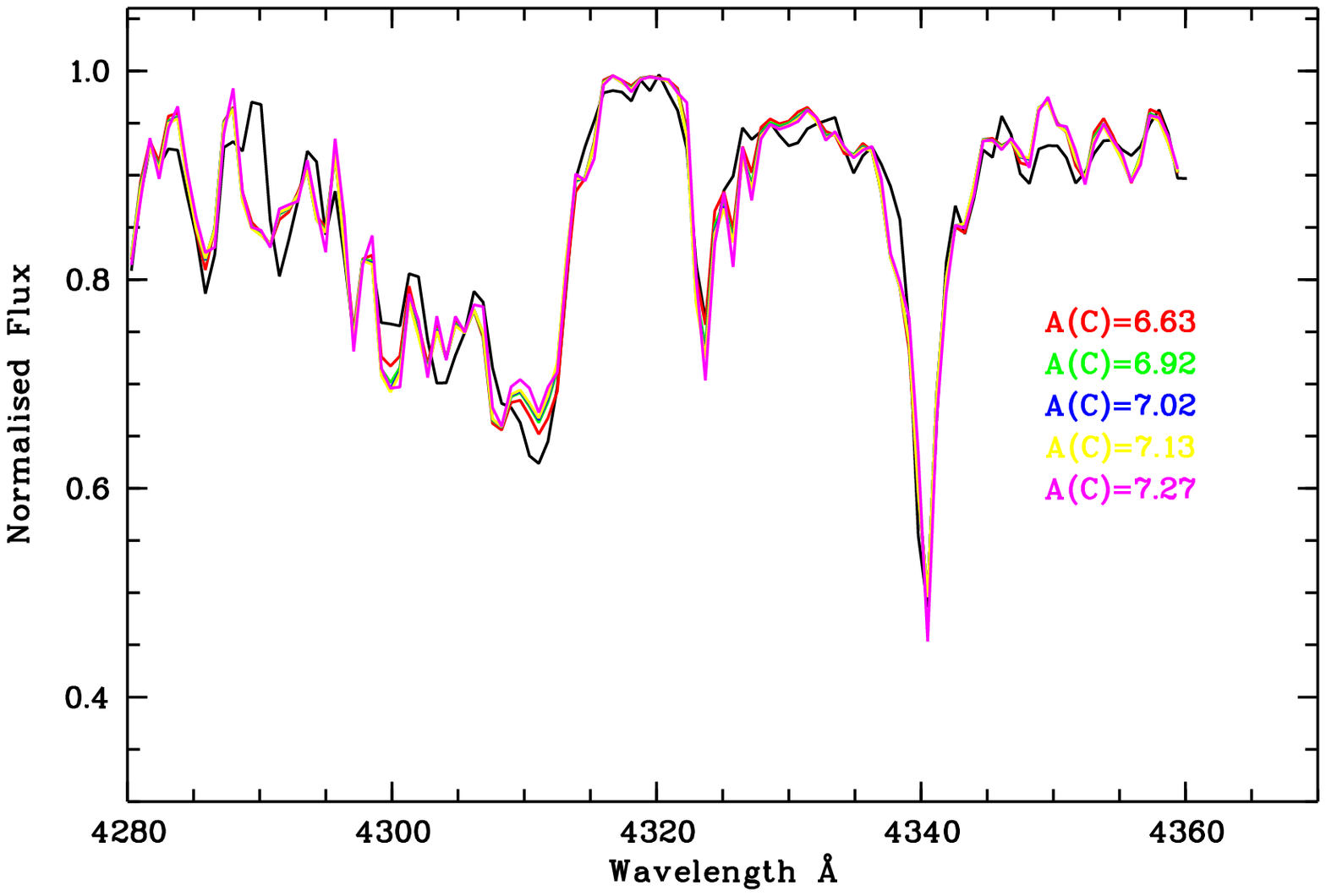}}
\end{center}
\caption[]{Observed spectrum of SDSS\,J1313-0010 (solid black)
  compared with the best fit based on a 3D synthesis with various C/O
  content. The pink profile corresponds to A(C)-A(O)=--1.67, and the
  red profile shows A(C)-A(O)=+0.33.}
\label{j1313gband}
\end{figure}

 The goal of the work presented in this paper is not to improve our knowledge of the chemical composition of this star with the FORS
  spectra. The aim is instead to test the 3D spectra and
  observationally verify if the C abundance we derive varies by
  changing the oxygen abundance simulated in the synthesis. The
  closest 3D model in the CIFIST grid has parameters \teff =5500\,K,
  \logg\ of 2.5\,[cgs], and [Fe/H]$=-3.0$; the model parameters are
  similar to those suggested by \citet{carlos15} (\Teff=5378\,K,
    \logg =3.0, [Fe/H]=-4.3, and [C/Fe]=+2.5]), with the exception of
the iron content. The 3D model was computed with a solar-scaled
  composition and an enhancement in $\alpha$-elements of 0.4\,dex,
  typical of a metal-poor star.  When computing the synthesis, several
  C/O ratios were considered. We then fitted the observed profile
  using a synthesis grid with the same C/O ratio \citep[see][for
    details]{andy17}. The fit was also made using a grid of
  1D LHD models \citep{zolfito}, computed with the same C/O ratios. At
  the typical temperatures, these stars have (5500-6500\,K), the
    A(C) value derived from the 1D fit is not sensitive to changing
    C/O: a negligible difference of 0.09\,dex is found when A(C)--A(O) is
    varied by up to 2.0. Conversely, the change in the C abundance from
    the 3D fit is 0.64\,dex when A(C)--A(O) is varied by the same
    amount. When the oxygen abundance is varied for any given carbon
    abundance, the shape of the {\it G}-band varies as well. This is most
    notable in the band head. However, a higher spectral
    resolution is required for this effect to be notable. The fits are shown in
  Fig.\,\ref{j1313gband}. The shape of the {\it G}-band also changes by
  changing C/O; the plot appears to favour a lower
  oxygen, and the $\chi^2$ test also favours this case.

\section{Results}

\begin{figure}
\begin{center}
\resizebox{\hsize}{!}{\includegraphics[draft = \draftflag, clip=true]{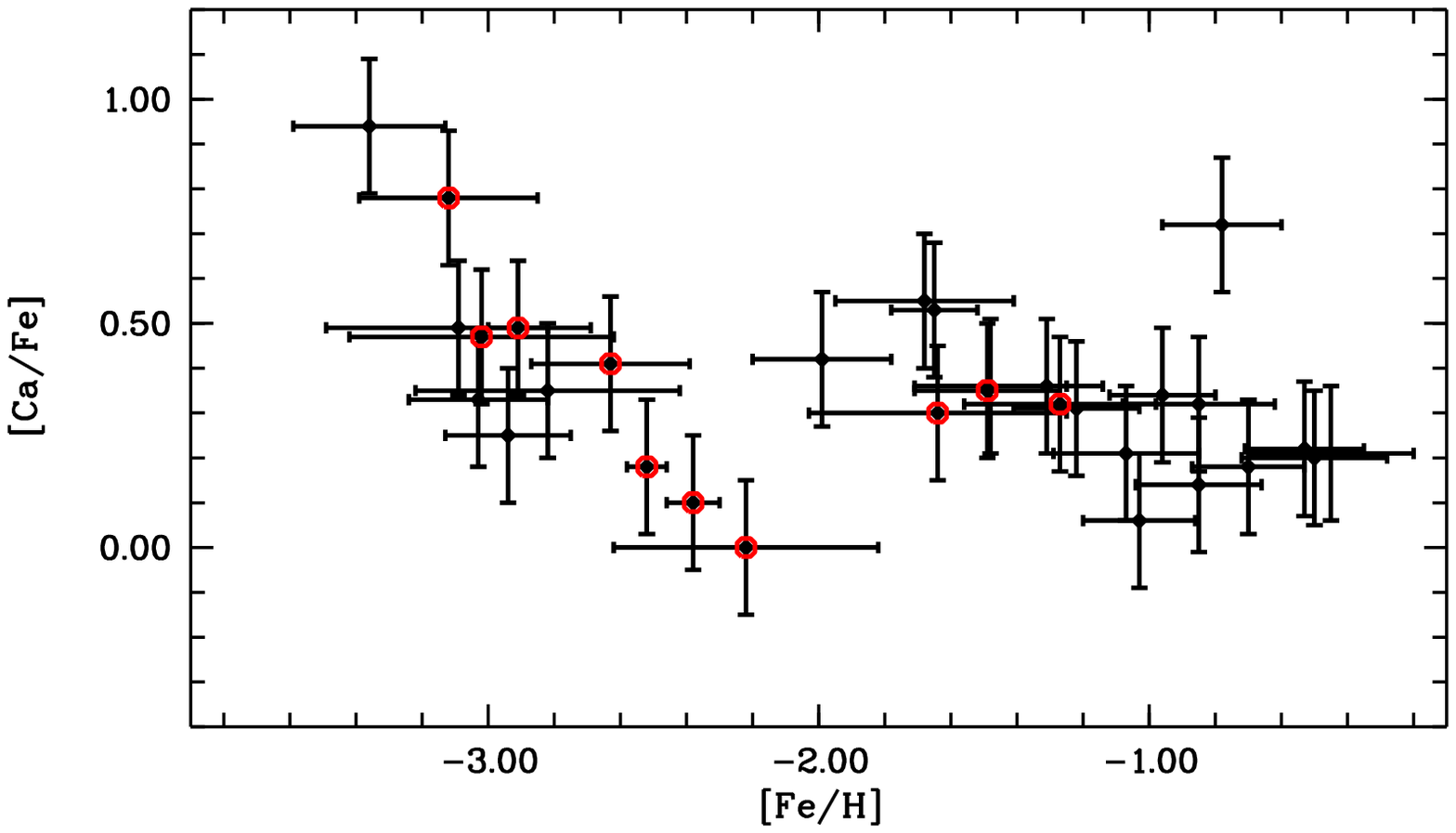}}
\end{center}
\caption[]{[Ca/Fe] derived from the \ion{Ca}{ii}-K line vs.
  [Fe/H].  The black dots surrounded by a red corona represent Ba-rich
  stars.  The solar A(Fe) value of 7.52 is taken from \citet{abbosun}, and the solar A(Ca)=6.33 is adopted from \citet{lodders09}.}
\label{cakfe}
\end{figure}

In Fig.\,\ref{cakfe} we show the [Ca/Fe] versus [Fe/H]. The Ca
abundance has been derived from the \ion{Ca}{ii}-K line for this
plot, for which we estimated an uncertainty of 0.15\,dex, while the
uncertainty on Fe is the line-to-line scatter, except for the stars
for which no clear Fe line is available; in this case, the uncertainty
is related to a comparison with synthetic spectra.  The stars, as
expected, all seem mostly $\alpha$-enhanced. The star with the lowest
[Ca/Fe] (SDSS\,J2219+0515) is one with the poorest A(Fe) determination,
so that within $1\,\sigma$ the star is a $\alpha$-enhanced metal-poor
star, at the same level as a typical metal-poor star.

\begin{figure}
\begin{center}
\resizebox{\hsize}{!}{\includegraphics[draft = \draftflag, clip=true]{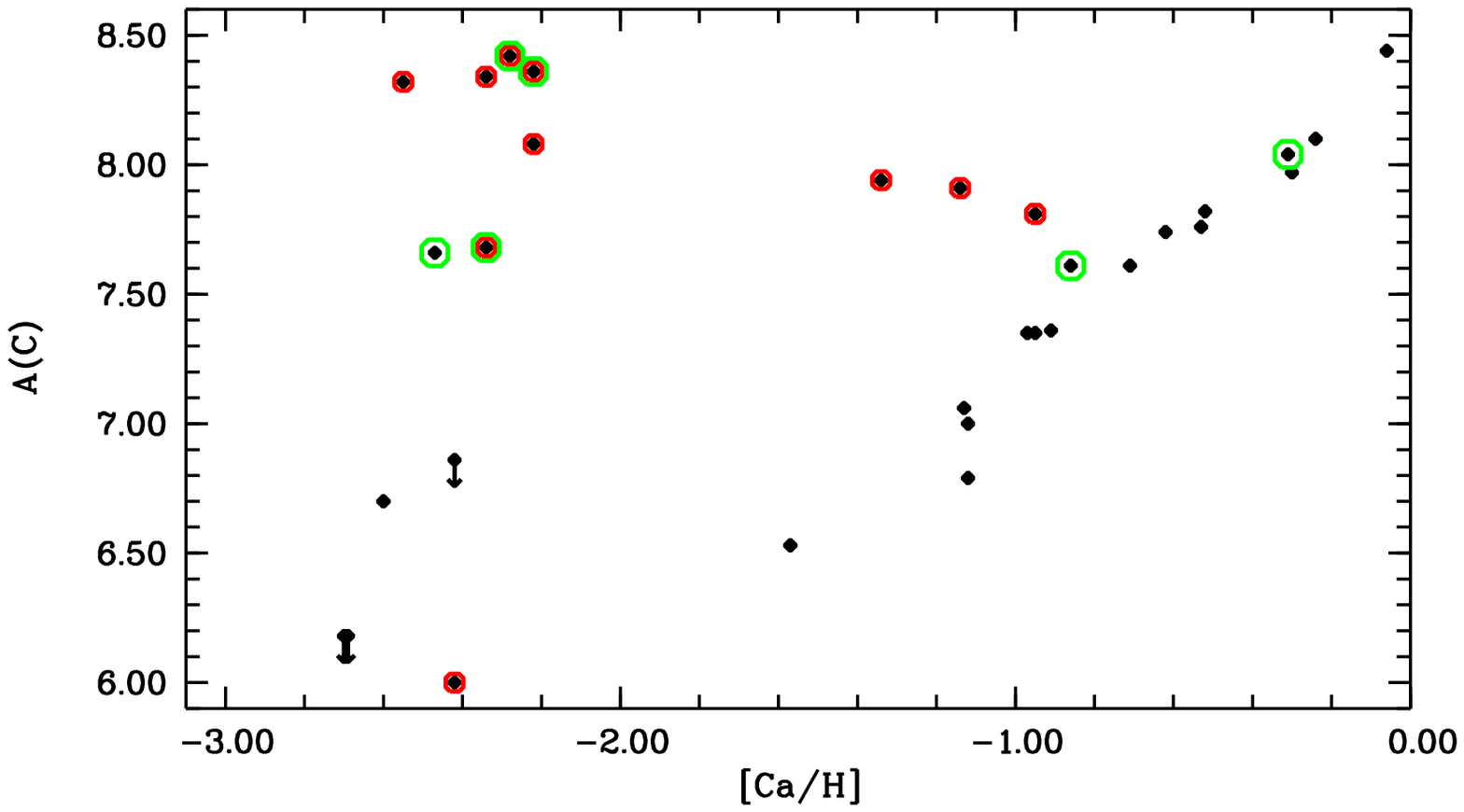}}
\end{center}
\caption[]{Carbon abundance vs. [Ca/H] derived from the \ion{Ca}{ii}-K
  line.  The black dots surrounded by a red corona represent Ba-rich stars.
  An extra green corona is used in the cases of uncertain measurements.
 Downward arrows represent upper limits.}
\label{accak}
\end{figure}

In Fig.\,\ref{accak} the carbon abundance, derived from the G
band and
listed in Table\,\ref{abbo}, is plotted as a function of the [Ca/H]
ratio, with the Ca abundance derived from the \ion{Ca}{ii}-K line. 
In the figure, the stars with $[{\rm Fe/H}]<-2$ are located in the higher part of the plot, which means that they
belong to the high C band of \citet{spite13} or have a low A(C)
value, placing them on the low C band.  Interestingly, all the stars in
the high C band are rich in Ba (two stars, SDSS\,J1543+0929 and
SDSS\,J2219+0515, have an uncertain A(Ba) value, which is depicted as a
  green corona in the figure).  The stars belonging to the low C band
are not rich in Ba (they have a low [Ba/Fe] or an upper limit): they are
CEMP-no stars.  Two stars (SDSS\,J0905+0330 and SDSS\,J1149+0723) with
a very similar A(C) value, just below 7.7, both belong to the higher
C band. The latter, SDSS\,J1149+0723, does not appear to be Ba-rich, but
the result, as for the other, is uncertain.

\begin{figure*}
\begin{center}
\resizebox{\hsize}{!}{\includegraphics[draft = \draftflag, clip=true]{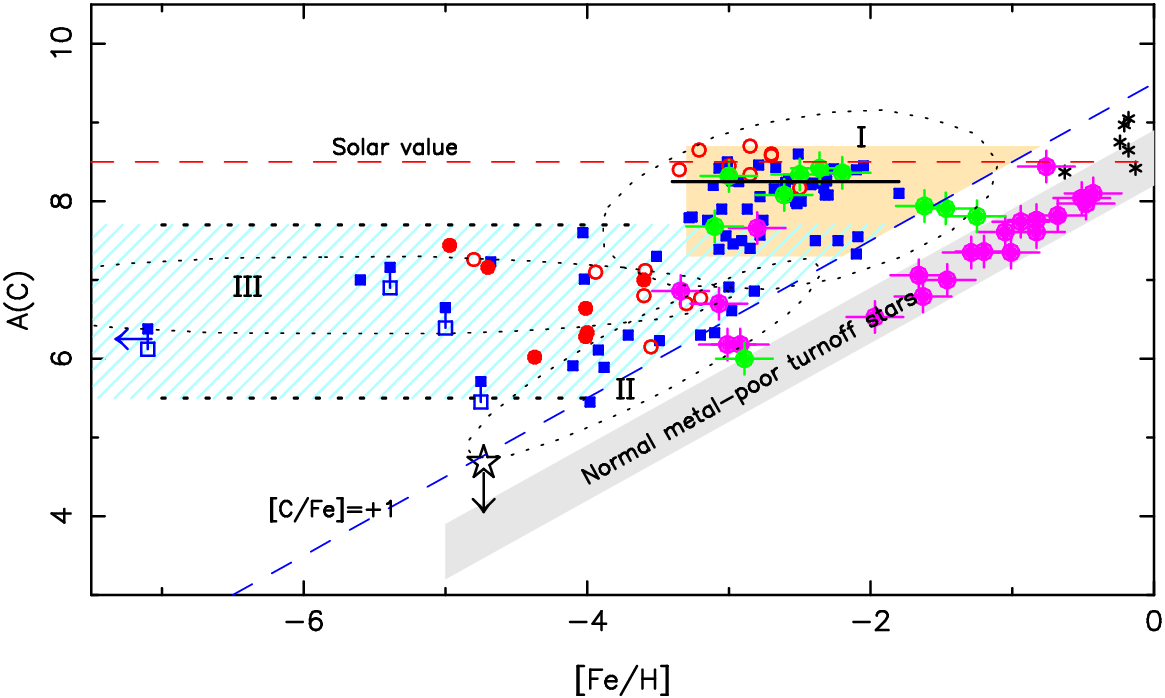}}
\end{center}
\caption[]{Carbon abundance vs. [Fe/H] of the sample stars
  (pink and green dots show stars normal and rich in Ba,
  respectively). Red solid dots are the stars we analysed in the TOPoS
  project \citep{topos2}; the open red dots are taken from
\citet{SivaraniBB06,BonifacioSC09,topos2,spite13,BeharaBL10,SivaraniBM04} and \citet{CaffauBF13,hydra};
  the black star is taken from \citet{stellina}. The black asterisks show known CH
  dwarf stars \citep{KarinkuzhiGos14,KarinkuzhiGos15}.  The blue
  squares show stars from the literature
  \citep{YongNB13,CohenCT13,CarolloFB14,MasseronJL12,Jonsell2006,ThompsonIB08,CohenCQ03,HansenHC15,HansenNH16,
    LucatelloGC03,Aoki2008,AokiRN02,2005Natur.434..871F,AokiFC06,LiZC15,NorrisCK07,ChristliebGK04,KellerBF14,frebel15,RoedererPT14,IvansSG05}.
 The blue and orange areas in the figure highlight the two carbon bands as defined in \citet{topos4}.
The black dashed ellipses represent the regions containing the three CEMP populations according to \citet{Yoon16}.}
\label{msacfeh}
\end{figure*}

We revisited Fig.\ref{accak} by comparing our sample of stars with the literature
in Fig.\,\ref{msacfeh}.
 This plot has been devised by \citet{spite13} and
  was recently revisited by \citet{topos4}, who also suggested a
  classification of CEMP stars according to C abundance different from the one
  of \citet{Yoon16}. The blue and orange areas in the figure highlight the two carbon bands as
defined in \citet{topos4}, and the three Yoon groups are delimited by dotted lines. 
\citet{Yoon16} adopted  [C/Fe]$>0.7$ instead of [C/Fe]$>+1.0$
for the definition of CEMP stars.
Moreover, in their Figure 1, \citet{Yoon16} also included giant stars. To define our two regions (orange and blue) 
in about the same sample of stars, we kept only dwarfs and turn-off stars where the carbon abundance 
is not affected by mixing. However, at very low metallicity ([Fe/H]$<-4.5$), we added four giant stars of 
the low red giant branch, but all with \logg\,$>2.2$. For these stars the correction of the carbon abundance 
due to the first dredge-up is small and has been estimated from \citet{BonifacioSC09}. 
As visible in Fig.\,\ref{msacfeh}, the two regions I and III defined in \citet{Yoon16} 
are included in our orange and blue areas \citep[as in][]{topos4}, respectively,
while stars in region II are either in our low C band (blue area) or are C-normal, according to our definition.
In our interpretation,
all C-enhanced stars ([C/Fe]$>+1.0$ as  defined  in the introduction) in the figure are adequately accounted for by the high and low C bands
\citep[as defined in][]{topos4}.
The stars analysed here are shown in Fig.\,\ref{msacfeh} as pink (normal in Ba) and green
(Ba rich) filled dots, in comparison with other stars taken from the literature.
The stars lying below the blue
dashed line are carbon-normal stars.  In this figure the CEMP ([Fe/H]$<-2.0$)
Ba-rich stars clearly belong to the high C band, while the stars with normal Ba
are part of the low C band. The two stars SDSS\,J0905+0330 and
SDSS\,J1149+0723, one normal and one rich in Ba, but with uncertain
measurements, are also similar in [Fe/H] and are placed in the lower
part of the high C band or in the upper part of the low C band.

\begin{figure}
\begin{center}
\resizebox{\hsize}{!}{\includegraphics[draft = \draftflag, clip=true]{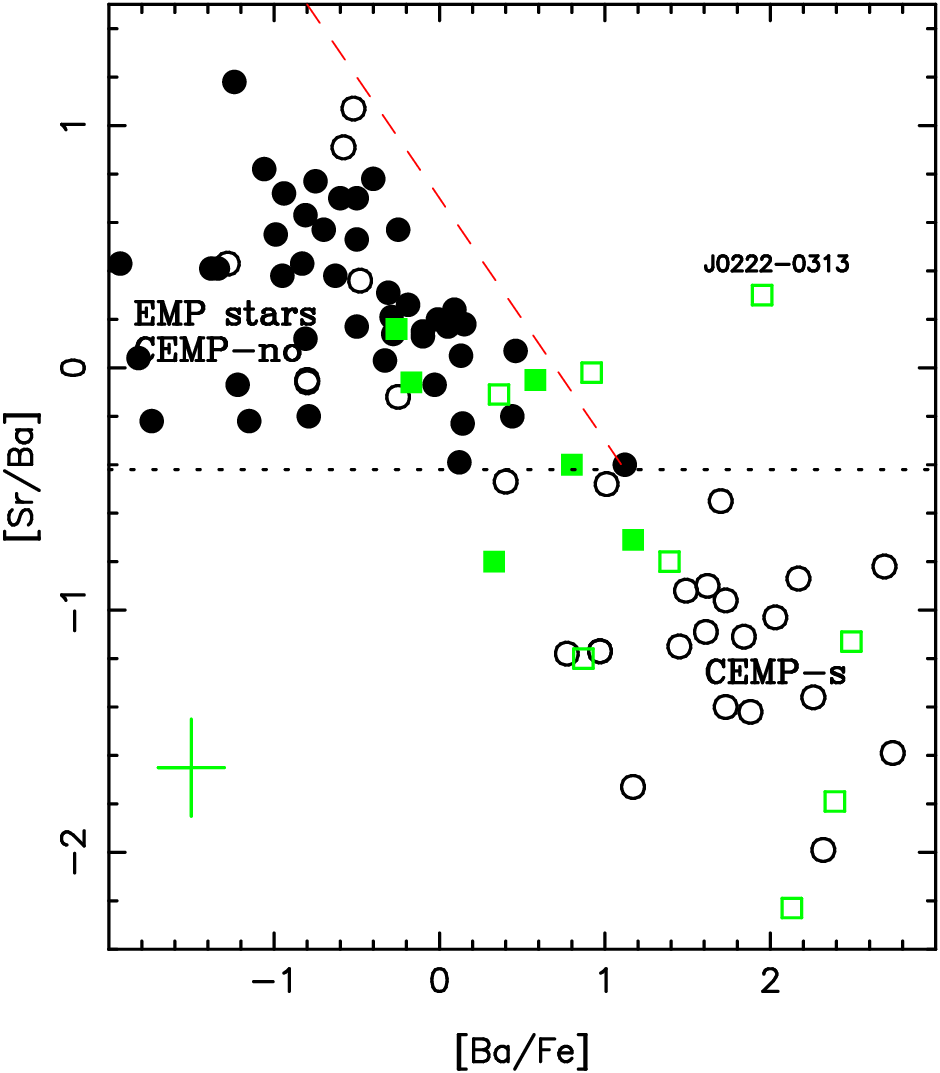}}
\end{center}
\caption[]{[Sr/Ba] ratio vs. [Ba/Fe].  The black filled circles
  represent the normal EMP dwarfs and giants studied in the frame of
  the ESO Large Programme First Stars \citep{CayrelDS04,BonifacioSC09}
  and the stars studied homogeneously by \citet{SiqueiraAB15}.  The
  black open circles are for CEMP stars studied in
  the frame of this ESO Large Programme or taken from the literature
  \citep{DepagneDS02,SivaraniBM04,SivaraniBB06,BarbuySS05,BeharaBL10,spite13,YongNB13}.
  The green filled squares show normal metal-poor stars studied in
  this paper, and the open green squares denote the C-rich stars.  Note
  the very peculiar position of the CEMP star SDSS\,J0222-0313. In
  this star the first-peak heavy elements (such as Sr) seem to be more
  enriched than the second-peak heavy elements (like Ba).}
\label{srba_bafe}
\end{figure}

In Fig.\ref{srba_bafe} we compare the position of the most
  metal-poor stars in our sample ([Fe/H]$<-1.5$) with a sample of
  metal-poor stars collected from the literature, which are, in
  general, slightly more metal poor than our sample. They are
  depicted in the form [Sr/Ba] versus [Ba/Fe].
These stars are generally slightly more metal poor than our sample
([Fe/H]$<-2.5$).  The ``normal'' stars (i.e. not C-rich) are
represented by filled symbols and the C-rich stars ([C/Fe]$>1.0$) by
open symbols. The green square symbols represent the stars
studied in this paper and the black circles show the stars taken 
  from the literature. In normal metal-poor stars, [Sr/Ba] is higher
than --0.4, a value observed in particular in the stars that
are strongly
enriched in neutron-capture elements, such as CS31082-001, which
is explained by a
pure r-process enrichment and is classified as ``r\,II''. The stars
SDSS\,J2123--0804 and SDSS\,J0042+0055 could also belong to this
class when we take the uncertainties into account.

In Fig.\,\ref{srba_bafe} several carbon-rich stars are located in the
same region as the ``normal'' metal-poor stars (CEMP-no).  Many
CEMP stars are enriched in heavy elements (CEMP-s or -rs), however,
such as in
general $\rm[Ba/Fe]>1$ and $\rm[Sr/Ba]<-0.4$; the second peak of
  the neutron-capture elements such as Ba is more enriched than the
  first-peak elements such as Sr.  SDSS\,J0222-0313 is characterised by
a high value of [Ba/Fe] ([Ba/Fe]=1.95) and an even higher value of
[Sr/Fe] ([Sr/Fe]=2.25).  The high abundance of Sr is also confirmed by
the Sr II line at 421.55\,nm. In this star the yttrium line (another
first-peak neutron-capture element) at 395.04\,nm was also
detected and measured: [Y/Fe]=+2.57.  It would be interesting to be
able to measure more neutron-capture elements in this star to explain
its unexpected abundance pattern.

\section{Conclusions}

 We analysed 30 unevolved stars and reinvestigated
  one known ultra Fe-poor star \citep{carlos15}.  In spite of the
  low resolution of the spectra, we were able to derive very useful
  information to better understand the C-enhanced stellar population.
  The CEMP stars belonging to the high C band are enhanced in Ba,
  while those belonging to the low band show a normal A(Ba) value.
  This finding supports the idea of \citet{topos2} that the high
  band is populated by binary stars and the high abundances are the
  result of mass-transfer from a companion, while the stars of the low
  band, with normal A(Ba), formed from a C-rich gas cloud.

We compared the [Sr/Ba] ratio as a function of [Ba/Fe] for the most
Fe-poor stars of this sample to a more metal-poor sample. In
Fig.~\ref{srba_bafe}, our stars appear to be similar to the most
metal-poor population; the Ba normal stars are mainly clustered in the upper left
side of the diagram and the Ba-rich stars lie in the lower right side.
One star, SDSS\,J0222--0313, is alone in the upper right part of the
plot. We need high-resolution observation for this star to confirm
with a higher degree of certainty that the first-peak elements (here
Sr and Y) are more abundant with respect to Fe than those of the
second peak (here Ba).
 
We also investigated the {\it G}-band of the known giant star
\citep{carlos15} using a synthesis computed from a hydrodynamical
model with parameters similar to those of the star.  Because
according to the 3D theoretical computation, the shape of the {\it G}-band
changes as a function of the C/O ratio, we might place constraints on
the oxygen abundance by fitting the {\it G}-band. In the case of
SDSS\,J1331--0019, a [O/Fe] lower than [C/Fe] is expected. Spectra with a higher resolution would be needed to confirm this, however. 


\begin{acknowledgements}
  This research has made use of the services of the ESO Science Archive
  Facility. AJG acknowledges the support of the Collaborative Research Centre
  SFB 881 (Heidelberg University) of the Deutsche Forschungsgemeinschaft (DFG,
  German Research Foundation).
  S.D. acknowledges support from Comit\'e Mixto ESO-GOBIERNO DE CHILE. 
\end{acknowledgements}

\bibliographystyle{aa}

\begin{thebibliography}{}

\bibitem[Allende Prieto et al.(2015)]{carlos15} Allende Prieto, C., Fern{\'a}ndez-Alvar, E., Aguado, D.~S., et al.\ 2015, \aap, 579, A98 

\bibitem[Alvarez \& Plez (1998)]{alvarez_plez} Alvarez R., Plez B., 1998, A\&A 330, 1109

\bibitem[Aoki et al.(2002)]{AokiRN02} Aoki, W., Ryan, S.~G., Norris, J.~E., et al.\ 2002, \apj, 580, 1149

\bibitem[Aoki et al.(2006)]{AokiFC06} Aoki, W., Frebel, A., Christlieb, N., et al.\ 2006, \apj, 639, 897

\bibitem[{{Aoki} {et~al.}(2007){Aoki}, {Beers}, {Christlieb}, {Norris}, {Ryan}, \& {Tsangarides}}]{Aoki2007} {Aoki}, W., {Beers}, T.~C., {Christlieb}, N., {et~al.} 2007, \apj, 655, 492

\bibitem[{{Aoki} {et~al.}(2008){Aoki}, {Beers}, {Sivarani}, {Marsteller}, {Lee}, {Honda}, {Norris}, {Ryan}, \& {Carollo}}]{Aoki2008} {Aoki}, W., {Beers}, T.~C., {Sivarani}, T., {et~al.} 2008, \apj, 678, 1351

\bibitem[Appenzeller et al.(1998)]{Appenzeller} Appenzeller, I., Fricke, K., F{\"u}rtig, W., et al.\ 1998, The Messenger, 94, 1 

\bibitem[Barbuy et al.(1997)]{Barbuy} Barbuy, B., Cayrel, R., Spite, M., et al.\ 1997, \aap, 317, L63 

\bibitem[Barbuy et al.(2005)]{BarbuySS05} Barbuy, B., Spite, M., Spite, F., et al.\ 2005, \aap, 429, 1031 

\bibitem[Beers et al.(1985)]{HKI} Beers, T.~C., Preston, G.~W., \& Shectman, S.~A.\ 1985, \aj, 90, 2089 

\bibitem[Beers et al.(1992)]{HKII} Beers, T.~C., Preston, G.~W., \& Shectman, S.~A.\ 1992, \aj, 103, 1987 

\bibitem[Beers \& Christlieb(2005)]{bc05} Beers, T.~C., \& Christlieb, N.\ 2005, \araa, 43, 531

\bibitem[Behara et al.(2010)]{BeharaBL10} Behara, N.~T., Bonifacio, P., Ludwig, H.-G., et al.\ 2010, \aap, 513, A72

\bibitem[Bidelman(1956)]{Bidelman} Bidelman, W.~P.\ 1956, Vistas in Astronomy, 2, 1428 

\bibitem[{{Bisterzo} {et~al.}(2006){Bisterzo}, {Gallino}, {Straniero}, {Ivans}, {K{\"a}ppeler}, \& {Aoki}}]{Bisterzo2006} {Bisterzo}, S., {Gallino}, R., {Straniero}, O., {et~al.} 2006, \memsai, 77, 985

\bibitem[Bonifacio et al.(1998)]{Bonifacio98} Bonifacio, P., Molaro, P., Beers, T.~C., \& Vladilo, G.\ 1998, \aap, 332, 672 

\bibitem[Bonifacio et al.(2009)]{BonifacioSC09} Bonifacio, P., et al.\ 2009, \aap, 501, 519 

\bibitem[Bonifacio et al.(2015)]{topos2} Bonifacio, P., Caffau, E., Spite, M., et al.\ 2015, \aap, 579, A28 

\bibitem[Bonifacio et al.(2018)]{topos4} Bonifacio, P., Caffau, E., Spite, M., et al.\ 2018, \aap, accepted, arXiv:1801.03935

\bibitem[Gaia Collaboration et al.(2016)]{gaiadr1} Gaia Collaboration, Brown, A.~G.~A., Vallenari, A., et al.\ 2016, \aap, 595, A2

\bibitem[Caffau \& Ludwig(2007)]{zolfito} Caffau, E., \& Ludwig, H.-G.\ 2007, \aap, 467, L11 

\bibitem[Caffau et al.(2011a)]{abbosun} Caffau, E., Ludwig, H.-G., Steffen, M., Freytag, B., \& Bonifacio, P.\ 2011a, \solphys, 268, 255

\bibitem[Caffau et al.(2012)]{stellina} Caffau, E., Bonifacio, P., Fran{\c c}ois, P., et al.\ 2012, \aap, 542, A51 

\bibitem[Caffau et al.(2013)]{CaffauBF13} Caffau, E., Bonifacio, P., Fran{\c c}ois, P., et al.\ 2013, \aap, 560, A15

\bibitem[Caffau et al.(2013)]{topos1} Caffau, E., Bonifacio, P., Sbordone, L., et al.\ 2013, \aap, 560, A71

\bibitem[Caffau et al.(2016)]{hydra} Caffau, E., Bonifacio, P., Spite, M., et al.\ 2016, \aap, 595, L6

\bibitem[Carollo et al.(2014)]{CarolloFB14} Carollo, D., Freeman, K., Beers, T.~C., et al.\ 2014, \apj, 788, 180

\bibitem[Cayrel et al.(2004)]{CayrelDS04} Cayrel, R., et al.\ 2004, \aap, 416, 1117

\bibitem[Christlieb(2003)]{Christlieb03} Christlieb, N.\ 2003, Reviews in Modern Astronomy, 16, 191 

\bibitem[Christlieb et al.(2004)]{ChristliebGK04} Christlieb, N., Gustafsson, B., Korn, A.~J., et al.\ 2004, \apj, 603, 708

\bibitem[Cohen et al.(2013)]{CohenCT13} Cohen, J.~G., Christlieb, N., Thompson, I., et al.\ 2013, \apj, 778, 56

\bibitem[Cohen et al.(2003)]{CohenCQ03} Cohen, J.~G., Christlieb, N., Qian, Y.-Z., \& Wasserburg, G.~J.\ 2003, \apj, 588, 1082

\bibitem[Cowan \& Rose(1977)]{cowanrose77} Cowan, J.~J., \& Rose, W.~K.\ 1977, \apj, 212, 149

\bibitem[Depagne et al.(2002)]{DepagneDS02} Depagne, E., Hill, V., Spite, M., et al.\ 2002, \aap, 390, 187 

\bibitem[Frebel et al.(2005)]{2005Natur.434..871F} Frebel, A., Aoki, W., Christlieb, N., et al.\ 2005, \nat, 434, 871

\bibitem[Frebel et al.(2015)]{frebel15} Frebel, A., Chiti, A., Ji, A.~P., Jacobson, H.~R., \& Placco, V.~M.\ 2015, \apjl, 810, L27

\bibitem[Gallagher et al.(2016)]{ajg16} Gallagher, A.~J., Caffau, E., Bonifacio, P., et al.\ 2016, \aap, 593, A48 

\bibitem[Gallagher et al.(2017)]{andy17} Gallagher, A.~J., Caffau, E., Bonifacio, P., et al.\ 2017, \aap, 598, L10

\bibitem[Gallagher et al.(2017)]{ajg17} Gallagher, A.~J., Steffen, M., Caffau, E., et al.\ 2017, \memsai, 88, 82 

\bibitem[Gimeno et al.(2016)]{GMOS_CCD} Gimeno, G., Roth, K., Chiboucas, K., et al.\ 2016, \procspie, 9908, 99082S 

\bibitem[Gustafsson et al. (2008)]{G2008} Gustafsson, B., Edvardsson, B., Eriksson, K., Graae-J{\o}rgensen, U., Nordlund, \AA., \& Plez, B.\ 2008, A\&A 486, 951

\bibitem[Hampel et al.(2016)]{hampel16} Hampel, M., Stancliffe, R.~J., Lugaro, M., \& Meyer, B.~S.\ 2016, \apj, 831, 171

\bibitem[Hansen et al.(2016)]{HansenNH16} Hansen, T.~T., Andersen, J., Nordstr{\"o}m, B., et al.\ 2016, \aap, 588, A3

\bibitem[Hansen et al.(2015)]{HansenHC15} Hansen, T., Hansen, C.~J., Christlieb, N., et al.\ 2015, \apj, 807, 173

\bibitem[Hook et al.(2004)]{GMOS} Hook, I.~M., J{\o}rgensen, I., Allington-Smith, J.~R., et al.\ 2004, \pasp, 116, 425 

\bibitem[Ivans et al.(2005)]{IvansSG05} Ivans, I.~I., Sneden, C., Gallino, R., Cowan, J.~J., \& Preston, G.~W.\ 2005, \apjl, 627, L145

\bibitem[{{Jonsell} {et~al.}(2006){Jonsell}, {Barklem}, {Gustafsson}, {Christlieb}, {Hill}, {Beers},
 \& {Holmberg}}]{Jonsell2006} {Jonsell}, K., {Barklem}, P.~S., {Gustafsson}, B., {et~al.} 2006, \aap, 451, 651

\bibitem[Karinkuzhi \& Goswami(2014)]{KarinkuzhiGos14} Karinkuzhi, D., \& Goswami, A.\ 2014, \mnras, 440, 1095

\bibitem[Karinkuzhi \& Goswami(2015)]{KarinkuzhiGos15} Karinkuzhi, D., \& Goswami, A.\ 2015, \mnras, 446, 2348

\bibitem[Keller et al.(2014)]{KellerBF14} Keller, S.~C., Bessell, M.~S., Frebel, A., et al.\ 2014, \nat, 506, 463

\bibitem[Li et al.(2015)]{LiZC15} Li, H.-N., Zhao, G., Christlieb, N., et al.\ 2015, \apj, 798, 110

\bibitem[Lodders et al.(2009)]{lodders09} Lodders, K., Plame, H., 
\& Gail, H.-P.\ 2009, Landolt-B{\"o}rnstein - Group VI Astronomy and Astrophysics
Numerical Data and Functional Relationships in Science and Technology Volume 
4B: Solar System.~ Edited by J.E.~Tr{\"u}mper, 2009, 4.4., 44 

\bibitem[Lucatello et al.(2003)]{Lucatello03} Lucatello, S., Gratton, R., Cohen, J.~G., et al.\ 2003, \aj, 125, 875 

\bibitem[Lucatello et al.(2005)]{lucatello05} Lucatello, S., Tsangarides, S., Beers, T.~C., et al.\ 2005, \apj, 625, 825

\bibitem[Lucatello et al.(2003)]{LucatelloGC03} Lucatello, S., Gratton, R., Cohen, J.~G., et al.\ 2003, \aj, 125, 875

\bibitem[Ludwig et al.(2009)]{ludwig09} Ludwig, H.-G., Caffau, E., Steffen, M., et al.\ 2009, \memsai, 80, 711

\bibitem[Masseron et al.(2012)]{MasseronJL12} Masseron, T., Johnson, J.~A., Lucatello, S., et al.\ 2012, \apj, 751, 14

\bibitem[McCarthy(1994)]{McCarthy} McCarthy, M.~F.\ 1994, The MK Process at 50 Years:  A Powerful Tool for Astrophysical Insight, 60, 224 

\bibitem[McClure(1984)]{McClure84} McClure, R.~D.\ 1984, \pasp, 96, 117 

\bibitem[McClure(1997)]{McClure97} McClure, R.~D.\ 1997, \pasp, 109, 536 

\bibitem[McClure \& Woodsworth(1990)]{MW90} McClure, R.~D., \& Woodsworth, A.~W.\ 1990, \apj, 352, 709 

\bibitem[McWilliam et al.(1995)]{McWilliam} McWilliam, A., Preston, G.~W., Sneden, C., \& Searle, L.\ 1995, \aj, 109, 2757 

\bibitem[Norris et al.(1997a)]{N97a} Norris, J.~E., Ryan, S.~G., \& Beers, T.~C.\ 1997a, \apj, 488, 350 

\bibitem[Norris et al.(1997b)]{N97b} Norris, J.~E., Ryan, S.~G., \& Beers, T.~C.\ 1997b, \apjl, 489, L169 

\bibitem[Norris et al.(2007)]{NorrisCK07} Norris, J.~E., Christlieb, N., Korn, A.~J., et al.\ 2007, \apj, 670, 774

\bibitem[{{Plez}(2012)}]{turbo}{Plez}, B. 2012, {Turbospectrum: Code for spectral synthesis}, astrophysics Source Code Library

\bibitem[Roederer et al.(2014)]{RoedererPT14} Roederer, I.~U., Preston, G.~W., Thompson, I.~B., Shectman, S.~A., \& Sneden, C.\ 2014, \apj, 784, 158

\bibitem[Sbordone et al.(2014)]{sbordone14} Sbordone, L., Caffau, E., Bonifacio, P., \& Duffau, S.\ 2014, \aap, 564, A109

\bibitem[Secchi(1868)]{Secchi} Secchi, A.\ 1868 Spettri Prismatici delle stelle fisse, Memoria Seconda, Rome, 1868, p. 11

\bibitem[Sivarani et al.(2004)]{SivaraniBM04} Sivarani, T., Bonifacio, P., Molaro, P., et al.\ 2004, \aap, 413, 1073

\bibitem[Sivarani et al.(2006)]{SivaraniBB06} Sivarani, T., Beers, T.~C., Bonifacio, P., et al.\ 2006, \aap, 459, 125 

\bibitem[Siqueira-Mello et al.(2015)]{SiqueiraAB15} Siqueira-Mello, C., Andrievsky, S.~M., Barbuy, B., et al.\ 2015, \aap, 584, A86

\bibitem[Sneden et al.(1996)]{Sneden} Sneden, C., McWilliam, A., Preston, G.~W., et al.\ 1996, \apj, 467, 819 

\bibitem[Spite et al.(2013)]{spite13} Spite, M., Caffau, E., Bonifacio, P., et al.\ 2013, \aap, 552, A107 

\bibitem[Starkenburg et al.(2014)]{Starkenburg} Starkenburg E., Shetrone M.~D., McConnachie A.~W., Venn K.~A., 2014, MNRAS, 441, 1217 

\bibitem[Troja et al.(2017)]{Troja2017} Troja, E., Piro, L.,
  van Eerten, H., et al.\ 2017, \nat, 551, 71 

\bibitem[Thompson et al.(2008)]{ThompsonIB08} Thompson, I.~B., Ivans, I.~I., Bisterzo, S., et al.\ 2008, \apj, 677, 556-571 

\bibitem[Wallerstein(1973)]{Wallerstein} Wallerstein, G.\ 1973, \araa, 11, 115 

\bibitem[Yanny et al.(2009)]{yanny09} Yanny, B., Rockosi, C., Newberg, H.~J., et al.\ 2009, \aj, 137, 4377

\bibitem[Yoon et al.(2016)]{Yoon16} Yoon, J., Beers, T.~C., Placco, V.~M., et al.\ 2016, \apj, 833, 20 

\bibitem[York et al.(2000)]{york00} York, D.~G., et al.\ 2000, \aj, 120, 1579

\bibitem[Yong et al.(2013)]{YongNB13} Yong, D., Norris, J.~E., Bessell, M.~S., et al.\ 2013, \apj, 762, 26

\end{thebibliography}

\appendix

\end{document}